\documentclass{aa}
\usepackage{graphicx}
\usepackage{amssymb}
\usepackage{amsmath}
\usepackage{natbib}
\usepackage{txfonts}
\usepackage{url}
\begin{document}
   \title{PHAT: PHoto-$z$ Accuracy Testing \thanks{Based on observations obtained with the Hubble Space Telescope, the Spitzer Space Telescope, the Keck Observatory, the Kitt Peak National Observatory, the Subaru Telescope, the Palomar Observatory, and the University of Hawaii 2.2-meter telescope.}}
%   \titlerunning{}

   \subtitle{}

   \author{H.~Hildebrandt\inst{1}
     \and S.~Arnouts\inst{2}
     \and P.~Capak\inst{3}
     \and L.~A.~Moustakas\inst{4}
     \and C.~Wolf\inst{5}
     \and F.~B.~Abdalla\inst{6}
     \and R.~J.~Assef\inst{7}
     \and M.~Banerji\inst{8}
     \and N.~Ben\'itez\inst{9}
     \and G.~B.~Brammer\inst{10}
     \and T.~Budav\'ari\inst{11}
     \and S.~Carliles\inst{12}
     \and D.~Coe\inst{4}
     \and T.~Dahlen\inst{13}
     \and R.~Feldmann\inst{14}
     \and D.~Gerdes\inst{15}
     \and B.~Gillis\inst{16}
     \and O.~Ilbert\inst{17}
     \and R.~Kotulla\inst{18,19}
     \and O.~Lahav\inst{6}
     \and I.~H.~Li\inst{20}
     \and J.-M.~Miralles\inst{21}
     \and N.~Purger\inst{22}
     \and S.~Schmidt\inst{23}
     \and J.~Singal\inst{24}
   }

   \offprints{H. Hildebrandt}

   \institute{Leiden Observatory, Leiden University, Niels Bohrweg 2, % 1
     2333CA Leiden, The Netherlands\\ \email{hendrik@strw.leidenuniv.nl} 
     \and 
     Canada-France-Hawaii Telescope Corporation, Kamuela, HI 96743, USA % 2
     \and
     Spitzer Science Center, 314-6, California Institute of % 3
     Technology, 1201 E.  California Blvd, Pasadena, CA, 91125, USA
     \and
     Jet Propulsion Laboratory, California Institute of Technology, MS 169-327, Pasadena, CA 91109, USA % 4
     \and
     Department of Physics, University of Oxford, DWB, Keble Road, Oxford, OX1 3RH, UK % 5
     \and
     Department of Physics and Astronomy, University College London, Gower Street, London WC1E 6BT, UK % 6
     \and
     Department of Astronomy, The Ohio State University, 4055 McPherson Lab, 140 W. 18th Avenue, Columbus, OH 43210, USA % 7
     \and
     Institute of Astronomy, University of Cambridge, Madingley Road, Cambridge, CB3 0HA, UK %8
     \and
     Instituto de Astrof\'isica de Andaluc\'ia (CSIC), Apdo. 3044, 18008 Granada, Spain
     \and
     Department of Astronomy, Yale University, New Haven, CT 06520-8101, USA % 10
     \and
     Department of Physics and Astronomy, Johns Hopkins University, 3400 North Charles Street, Baltimore, MD 21218, USA % 11
     \and
     Department of Computer Science, Johns Hopkins University, 3400 North Charles Street, Baltimore, MD 21218, USA % 12
     \and
     Space Telescope Science Institute, 3700 San Martin Drive, Baltimore, MD 21218, USA % 13
     \and
     Department of Physics, Institute of Astronomy, ETH Z\"urich, Wolfgang-Pauli-Strasse 16, 8093 Z\"urich, Switzerland % 14
     \and
     Department of Physics, University of Michigan, Ann Arbor, Michigan 48109, USA % 15
     \and
     Department of Physics and Astronomy, University of Waterloo, 200 University Avenue West, Waterloo, Ontario, N2L 3G1, Canada % 16
     \and
     Laboratoire d'Astrophysique de Marseille, CNRS-Universit\'e d'Aix-Marseille, 38 rue Fr\'ed\'eric Joliot-Curie, 13388 Marseille Cedex 13, France % 17
     \and
     Centre for Astrophysics Research, University of Hertfordshire, College Lane, Hatfield AL10 9AB, UK % 18
     \and
     Department of Astronomy, University of Wisconsin-Madison, 475 N Charter St, Madison, WI 53706, USA % 19
     \and
     Centre for Astrophysics \& Supercomputing, Swinburne University of Technology, PO Box 218, Hawthorn, VIC 3122, Australia %20
     \and
     Institut d'Estudis Andorrans, Avda Rocafort 21-23, AD600 Sant Juli\`a de L\`oria, Andorra % 21
     \and
     Department of Physics of Complex Systems, E\"otv\"os Lor\'and University, Pf. 32, H-1518 Budapest, Hungary % 22
     \and
     Physics Department, University of California, 1 Shields Avenue, Davis, CA 95616, USA % 23
     \and
     Kavli Institute for Particle Astrophysics and Cosmology, SLAC National Accelerator Laboratory, Menlo Park, CA 94025, USA % 24
     }
   \date{Received ; accepted }

  \abstract {Photometric redshifts (photo-$z$'s) have become an
    essential tool in extragalactic astronomy. Many current and
    upcoming observing programmes require great accuracy of photo-$z$'s
    to reach their scientific goals.}{Here we introduce PHAT, the
    PHoto-$z$ Accuracy Testing programme, an international initiative to
    test and compare different methods of photo-$z$ estimation.}{Two
    different test environments are set up, one (PHAT0) based on
    simulations to test the basic functionality of the different
    photo-$z$ codes, and another one (PHAT1) based on data from the
    GOODS survey including 18-band photometry and $\sim2\,000$
    spectroscopic redshifts.}{The accuracy of the different methods is
    expressed and ranked by the global photo-$z$ bias, scatter, and
    outlier rates. While most methods agree very well on PHAT0 there
    are differences in the handling of the Lyman-$\alpha$ forest for
    higher redshifts. Furthermore, different methods produce photo-$z$
    scatters that can differ by up to a factor of two even in this
    idealised case. A larger spread in accuracy is found for
    PHAT1. Few methods benefit from the addition of mid-IR
    photometry. The accuracy of the other methods is unaffected or
    suffers when IRAC data are included. Remaining biases and
    systematic effects can be explained by shortcomings in the
    different template sets (especially in the mid-IR) and the use of
    priors on the one hand and an insufficient training set on the
    other hand. Some strategies to overcome these problems are
    identified by comparing the methods in detail. Scatters of 4-8\%
    in $\Delta z/(1+z)$ were obtained, consistent with other
    studies. However, somewhat larger outlier rates ($>7.5\%$ with
    $\Delta z/(1+z)>0.15$; $>4.5\%$ after cleaning) are found for all
    codes that can only partly be explained by AGN or issues in the
    photometry or the spec-$z$ catalogue. Some outliers were probably
    missed in comparisons of photo-$z$'s to other, less complete
    spectroscopic surveys in the past. There is a general trend that
    empirical codes produce smaller biases than template-based
    codes.}{The systematic, quantitative comparison of different
    photo-$z$ codes presented here is a snapshot of the current
    state-of-the-art of photo-$z$ estimation and sets a standard for
    the assessment of photo-$z$ accuracy in the future. The rather
    large outlier rates reported here for PHAT1 on real data should be
    investigated further since they are most probably also present
    (and possibly hidden) in many other studies. The test data sets
    are publicly available and can be used to compare new, upcoming
    methods to established ones and help in guiding future photo-$z$
    method development.}


   \maketitle
%
%________________________________________________________________

\section{Introduction}
The estimation of redshifts from photometry alone is an old idea
\citep{1962IAUS...15..390B,1982ApJ...257L..57P,1985AJ.....90..418K,1986ApJ...303..154L,1995AJ....110.2655C}. It
has come a long way from being a rarely used technique for special
kinds of objects to a major tool now widely used for a multitude of
observational programmes.

Not only can this photometric redshift (photo-$z$) approach yield
redshifts of fainter objects than accessible by spectroscopy, but also
the efficiency in terms of the number of objects with redshift
estimates per unit telescope time is largely increased. These two
properties make photo-$z$'s extremely attractive for observing
programmes depending on redshifts for a large number of faint galaxies
if these redshifts do not have to be as precise as spectroscopic
redshifts (spec-$z$'s).

Still the requirements on the accuracy of photo-$z$'s for upcoming
surveys are formidable. Photo-$z$'s are essential in constraining dark
energy (DE) by weak gravitational lensing and can be used for other DE
probes such as galaxy clustering, supernovae of type Ia, and the mass
function of galaxy clusters as well
\citep{2006astro.ph..9591A,2006ewg3.rept.....P}. Surveys of galaxy
formation and evolution also depend on photo-$z$'s to study these
processes as a function of environment and to probe to fainter levels
than with spectroscopy alone. To fully exploit the power of these
huge, future data sets, photo-$z$'s with a very low level of residual
systematics are needed \citep[e.g. ][]{2006MNRAS.366..101H}.

There are many aspects which influence the performance of
photo-$z$'s. The choice of an observing strategy sets the theoretical
limit for the accuracy. Choosing the filters and distributing the
available observing time over the different filters to reach certain
depths can have a great impact on photo-$z$'s. Accurate photometric
calibration is of great importance as well as the removal of effects
of the different point-spread-function (PSF) in the different
bands. Varying column densities of galactic dust over the survey area
have to be accounted for before a photo-$z$ code can be expected to
perform at its best.

Here we would like to ignore all these effects as much as possible and
concentrate on the last link in the chain, the photo-$z$ methods
themselves. It is clear that the two regimes -- data and method --
cannot be separated cleanly because there are connections between the
two. For example, it is highly likely that one method of photo-$z$
estimation will perform better than a second method on one particular
data set while the situation may well be reversed on a different data
set. Whenever such a situation arises in the following we will try to
alert the reader to that.

The methodology behind photo-$z$'s is developing fast with ever more
complex methods yielding results of increasing accuracy. In this
context it is important to set a standard to compare the different
methods to each other in order to make quantitative statements about
their differences and to take a snapshot of today's
state-of-the-art. Such comparisons and rankings can then be used to
identify the most promising approaches and to concentrate on their
further improvement.

In this paper we present an international initiative named PHAT
(PHoto-$z$ Accuracy
Testing)\footnote{\url{http://www.astro.caltech.edu/twiki_phat/bin/view/Main/WebHome}}
which was initiated to carry out such a quantitative comparison. A
very similar initiative has been carried out for shape measurement
algorithms in the Shear TEsting Program \citep[STEP;
][]{2006MNRAS.368.1323H,2007MNRAS.376...13M} and led to important
improvements in the methodology of measuring galaxy shapes for weak
gravitational lensing applications. Similar but much more limited
blind tests of photo-$z$'s have been performed by
\cite{1998AJ....115.1418H} on spectroscopic data from the Keck
telescope on the Hubble Deep Field (HDF), by
\cite{2008A&A...480..703H} on spectroscopic data from the VIMOS VLT
Deep Survey \citep[VVDS; ][]{2004A&A...428.1043L} and the FORS Deep
Field \citep[FDF; ][]{2004A&A...418..885N}, and by
\cite{2008arXiv0812.3831A} on the sample of Luminous Red Galaxies from
the SDSS-DR6.

In the framework of PHAT we provide standardised test environments to
the photo-$z$ community which consist of simulated or observed
photometric catalogues alongside with additional material like filter
curves, SED templates, and training sets. These data sets can be used
in a blind (or semi-blind, i.e. with support of a training set) test
by the participants to estimate redshifts with their favourite
codes. Two such test steps have been carried out so far. The first one
called PHAT0 is based on a highly idealised simulation representing an
easy case to test the most basic elements of photo-$z$ estimation and
to identify possible low-level discrepancies between the methods. The
second test called PHAT1 is based on real data originating from the
Great Observatories Origins Deep Survey \citep[GOODS,
][]{2004ApJ...600L..93G} representing a much more complex environment
pushing photo-$z$ codes to their limits and revealing more systematic
difficulties.

PHAT was conceived as an open competition. The test data sets are
publicly available over the PHAT website and all major photo-$z$
groups in the astronomical community were informed of the initiative
via email. Furthermore, PHAT was advertised on several meetings and
workshops to increase its visibility. The photo-$z$ codes presented
here were not selected by the PHAT coordinators but reflect the
interest of the community in such a competition. This strategy led to
an impressive feedback of 21 participants submitting results obtained
with 17 different photo-$z$ codes. After a large number of results was
collected for each test data set, the results of all codes were
published on the PHAT website. But the test data sets are still kept
blind (i.e. the individual redshifts are retained) to allow further
participants to meet the same conditions.

First we shortly summarise every photo-$z$ method that was used
within PHAT (Sect.~\ref{sec:methods}). Then in
Sect.~\ref{sec:PHAT0}~\&~\ref{sec:PHAT1} the motivation behind the
tests, the data sets, and the results are described in detail for
PHAT0 and PHAT1, respectively. In Sect.~\ref{sec:conclusions} we
conclude and give an outlook to future activities within PHAT. We use
AB magnitudes throughout.

\section{Methods}
\label{sec:methods}
In the following we describe the different methods that were used to
estimate photo-$z$'s from the catalogues presented in
Sect.~\ref{sec:PHAT0}~\&~\ref{sec:PHAT1}. A summary of the methods can
also be found in Table~\ref{tab:methods} together with the
three-letter acronyms that are used in the remainder of the paper to
identify the codes. The third small letter indicates whether the code
belongs to the empirical codes (-e), which are trained on the colours
\footnote{Most empirical codes offer the flexibility of using also any
  other photometric observable like e.g. size, concentration, or
  surface brightness. Since we only use magnitudes in PHAT we skip
  this detail in the remainder of Sect.~\ref{sec:methods}.} of a
sub-sample of objects with accurate redshift estimates
(e.g. spec-$z$'s), or to the codes fitting SED templates to the
observed photometry (-t). It should be noted that this distinction is
somewhat fuzzy. A number of codes include ingredients from both
regimes. We just keep this terminology because it has been widely used
in the literature. For a more rigorous description of the underlying
concepts in photo-$z$ methods and their common properties see
\cite{2009ApJ...695..747B}.

Note that the descriptions of the different template sets of the
template SED fitting codes in the following subsections only apply to
PHAT1. For PHAT0 the template set was provided and it was used by
every participant with a template-based code.

\subsection{\texttt{BPZ} (BP-t)}
\texttt{BPZ} \citep[Bayesian Photo-$z$'s;][]{2000ApJ...536..571B}
introduced the use of Bayesian inference and priors to photometric
redshift estimation. The code uses a prior $P(z,T \mid m_0)$ which
gives the likelihood that given an apparent magnitude $m_0$, a galaxy
would have redshift $z$ and SED type $T$. As an example of how the
prior works, bright objects and ellipticals are assumed unlikely to be
at high redshift.  For each galaxy, this information is combined (in a
Bayesian manner) with the likelihood $P(C \mid z,T)$ of observing the
galaxy colours $C$ for each redshift and SED pair, yielding the final
$P(z,T \mid C,m_0)$.  By marginalising over $T$, $P(z)$ is obtained
along with the most likely redshift $z_b$ and its uncertainties. For
the PHAT tests, \texttt{BPZ} version 1.99.3 is used, a slightly
updated version of that used in the \cite{2006AJ....132..926C} UDF
analysis.

\begin{itemize}
\item {\it Templates:} The \cite{2006AJ....132..926C} SED templates
  are used with \texttt{BPZ}, which include a CWW+SB SED template set
  (similar to that used in PHAT0 with Kinney et al.'s SB1 replaced by
  SB3) as introduced in \cite{2000ApJ...536..571B} and re-calibrated
  by \cite{2004ApJS..150....1B} plus two younger starburst templates
  from \cite{2003MNRAS.344.1000B} added in \cite{2006AJ....132..926C}.
  Note that the empirical CWW+SB templates as well as the synthetic
  BC03 templates include emission lines. No dust extinction was added
  to the BC03 templates. Between each of the eight adjacent templates
  two interpolated templates are added, for a total of 22 templates.
  Beyond $25600\AA$, the majority of the templates are undefined and
  must be extrapolated.  Thus it cannot be expected that these
  templates provide good fits to IRAC photometry of low redshift
  objects.
\item {\it Prior:} For PHAT0, a flat prior is used. The prior was
  calculated by \cite{2000ApJ...536..571B} based on objects with
  spec-$z$ in the CFRS \citep{1995ApJ...455...50L} and HDF-N
  \citep{1996AJ....112.1335W}.  It was shown to yield results superior
  to the ``flat'' prior implicitly assumed by maximum likelihood (or
  ``frequentist'') methods.
\item {\it Training:} No training with the model-$z$'s/spec-$z$'s was
  performed.
\end{itemize}

\subsection{\texttt{BPZ} (BP2-t)}
\texttt{BPZ} is run on PHAT1 a second time with a different template
set and additional training.

\begin{itemize}
\item {\it Templates:} The second library (Ben\'itez 2010, in
  preparation) uses as starting point a set of 6 templates from PEGASE
  \citep{1997A&A...326..950F} selected to be similar to the
  \cite{2006AJ....132..926C} templates. This library is further
  calibrated using the FIREWORKS photometry and spectroscopic
  redshifts from \cite{2008ApJ...682..985W}. Note that these
    templates include emission lines and dust extinction.
\item {\it Prior:} Same as BP-t.
\item {\it Training:} The templates are compared to the photometry of
  the spec-$z$ training set and new zero points are estimated, as in
  \cite{2006AJ....132..926C}. We also measure the amount of excess
  scatter in the predicted vs measured colours compared with that
  expected from the catalogue photometric errors and typical template
  uncertainties \citep{2008ApJ...686.1503B}. This excess scatter is
  included in the photo-$z$ estimation as a zero point uncertainty.
\end{itemize}

\subsection{\texttt{EAZY} (EA-t)}
\texttt{EAZY} \citep{2008ApJ...686.1503B} is a template-fitting code
designed to produce un-biased photometric redshift estimates for deep
multi-wavelength surveys that lack representative calibration samples
with spectroscopic redshifts.  

\begin{itemize}
\item {\it Templates:} \texttt{EAZY} uses a unique template set
  derived using the non-negative matrix factorisation algorithm
  \citep{nmf, blanton:07} trained on synthetic photometry from the
  semi-analytic light-cone produced by \cite{delucia:07}.  These
  templates can be considered the principal component spectra of all
  galaxies at $0<z<4$ in the light-cone, allowing for subtle
  differences between local and high-redshift galaxy samples.
  \texttt{EAZY} is able to reproduce complex star-formation histories
  by fitting non-negative linear combinations of the templates.  The
  templates include emission lines following the prescription of
  \cite{2009ApJ...690.1236I}.
\item {\it Template error function:} Template mismatch is addressed
  with a ``template error function'', which assigns lower weights at
  rest-frame wavelengths where the template calibration is uncertain
  or where the templates are not expected to fully reproduce observed
  galaxy colours.  This feature is particularly important when using
  mid-IR (IRAC) photometry, which samples wavelengths where the
  observed emission can be dominated by non-stellar (i.e.  dust)
  sources not included in the templates.
\item {\it Prior:} EAZY adopts a prior equal to the normalised
  redshift distribution of galaxies in the \cite{delucia:07}
  semi-analytic light-cone at a given apparent $R$ or $K$ magnitude.
  This is akin to a luminosity prior under the assumption that the
  light-cone reasonably reproduces the galaxy luminosity function.
\item {\it Training:} No training with the model-$z$'s/spec-$z$'s was
  performed.
\end{itemize}

\subsection{\texttt{GALEV} and \texttt{GAZELLE} (GA-t)}
\texttt{GAZELLE} \citep[Kotulla, in preparation]{KotullaFritze09} is
based on a $\chi^2$ minimisation algorithm to compare the observed
SEDs to a large library of \texttt{GALEV} evolutionary synthesis
models \citep{2009MNRAS.396..462K}. \texttt{GAZELLE} also accounts for
inherent uncertainties in the model grid, e.g. due to uncertainties in
the stellar evolution data and stellar spectral libraries, by assuming
a $0.1$ mag uncertainty in all filters.

\begin{itemize}
\item {\it Templates:} \texttt{GALEV} includes a full suite of
  emission lines \citep{AndersFritze03}, a detailed treatment of the
  attenuation due to intergalactic HI \citep{1995ApJ...441...18M} and
  optionally a chemical evolution model. This combination allows to
  not only estimate photometric redshifts, but at the same time
  physical parameters (stellar masses, star formation rates, etc.) for
  each galaxy in a consistent manner. Masses and mass-dependent
  parameters are computed by scaling model values with the scaling
  factor derived from matching the overall normalisation of the
  template fluxes relative to the observed fluxes. For the PHAT1 run
  the model grid included 5 undisturbed models for E and Sa-Sd type
  galaxies supplemented with a set of 21 models encountering a strong
  starburst at galaxy ages of $0.5$ to $10$ Gyrs, followed by
  subsequent post-starburst phases. All models assume star formation
  to begin at $\rm z=8$; for the undisturbed models a chemically
  consistent evolution (see \citealt{KotullaFritze09} for details) is
  chosen, for the burst models a metallicity fixed to half the solar
  value is used. All templates include the full evolution from the
  onset of star formation until the present day and the
  \cite{2000ApJ...533..682C} dust extinction description is
  chosen. Emission lines are included as well.
\item {\it Filter weighting:} To avoid complications at wavelengths
  beyond the rest-frame K-band where dust emission becomes
  increasingly important, only filters that cover the rest-frame
  K-band or shorter wavelengths are included, effectively ignoring
  some of the Spitzer filters at low-redshift.
\item {\it Prior:} No prior is included that might affect the
  resulting redshift distribution.
\item {\it Training:} No training with the model-$z$'s/spec-$z$'s was
  performed.
\end{itemize}

\subsection{\texttt{GOODZ} (GO-t)}
The \texttt{GOODZ} code (Dahlen et al. 2010, in preparation) is a
developed version of the code used by
\cite{2005ApJ...631..126D,2007ApJ...654..172D} to calculate
photometric redshifts in the GOODS-S. The code is based on the
template fitting method and allows the inclusion of Bayesian priors
based on the expected shape of the galaxy luminosity function.
Similar to this investigation, \texttt{GOODZ} uses the four empirical
templates from \cite{1980ApJS...43..393C} and two templates from
\cite[][their templates SB2 and SB3]{1996ApJ...467...38K}. The code
also uses available spectroscopic redshifts to correct for offsets
between fluxes extracted in different filters or instruments. Such
offsets may be significant when combining data from different
instruments with varying PSF or pixel-scales and may uncorrected lead
to increased scatter or biases in the photometric redshifts. The
spectroscopic redshifts are also used to adjust the input set of
template SEDs using a method similar to \cite{2006A&A...457..841I}.

\begin{itemize}
\item {\it Templates:} \texttt{GOODZ} is only run on PHAT0 so that no
  individual template set is associated with this code.
\item {\it Prior:} No prior was used.
\item {\it Training:} No training with the model-$z$'s was performed.
\end{itemize}

\subsection{\texttt{Hyperz} (HY-t)}
\texttt{Hyperz} is a publicly available code based on SED templates
fitting using a standard $\chi^2$ minimisation method. The codes uses
the observed fluxes of an object in a set of given filters and
compares them with the theoretical fluxes of galaxies in the same
filters obtained from template spectra, either synthetic or empirical,
taking into account the observational uncertainties but also the
possible observational hidden effects such as reddening or IGM
opacity. It computes not only a best-fit solution which minimises the
differences, therefore a most probable photometric redshift, but also
a full probability function as a function of redshift. The code and
the method have been tested and described extensively in
\cite{2000A&A...363..476B} and further practical description can be
found in its users manual. \texttt{Hyperz} comes with a given set of
templates, filters, reddening laws and Lyman forest modelling but can
be easily adapted to use any kind of parameters that would fit the
needs of the user. Its simplicity has brought \texttt{Hyperz} to be
extensively used and tested since its launch, and even to be used
beyond the pure computation of photometric redshifts.

\begin{itemize}
\item {\it Templates:} \texttt{Hyperz} comes with two standard
  template sets, one based on the synthetic stellar population library
  of \cite{1993ApJ...405..538B} and the other one consisting of the
  four empirical templates from \cite{1980ApJS...43..393C}. For the
  PHAT1 test, the latter empirical library was chosen and it was
  supplemented with two starburst templates from
  \cite{1996ApJ...467...38K} (templates from both libraries include
  emission lines). This set of six basic template was further enlarged
  by applying different amounts of extinction to the templates
  according to the \cite{2000ApJ...533..682C} dust extinction law.
\item {\it Prior:} No prior was used.
\item {\it Training:} No training with the model-$z$'s/spec-$z$'s was
  performed.
\end{itemize}

\subsection{\texttt{Kernelz} (KR-t)}
This method is a hybrid incorporating aspects of both template-based
and empirical codes, though it is most similar in design to
\texttt{BPZ} and other Bayesian methods. As in standard template-based
codes model colours are computed for a set of galaxy SEDs at a set of
fixed redshifts.  However, then this grid of colours is treated as if
they were individual galaxies. For each test galaxy the points are
weighted by a factor that is akin to a Bayesian prior, accounting for
the expected probability of seeing such a galaxy given the apparent
magnitude and type of the test point.  Redshifts are then estimated
using kernel regression, constructing a weighted average redshift,
with weights proportional to their proximity to the template points in
colour space. The kernel bandwidth is chosen by cross-validation using
the training set of galaxies with known redshifts.  Results presented
here represent code that is still in development.  Details of the
kernel regression method for both empirical and hybrid techniques will
be described in Schmidt \& Brewer (in prep).  A promising extension
that improves the method by allowing for data adaptive kernels will be
described in Udaltsova \& Schmidt (in prep).  A public release of the
code is also in the works.

\begin{itemize}
\item {\it Templates:} Because \texttt{Kernelz} was still in
  development when the results were submitted, simple templates from
  \cite{1980ApJS...43..393C} and \cite{1996ApJ...467...38K} (both of
  which include emission lines) with some extrapolation to IRAC
  wavelengths were used.
\item {\it Prior:} An empirical prior trained on data from VVDS was
  used.  In practice, this is very similar to the prior described in
  \cite{2006A&A...457..841I}.
\item {\it Training:} The spectroscopic data was used to choose the
  kernel bandwidth alone, no tweaking of templates or zero points was
  performed.
\end{itemize}

\subsection{\texttt{Le~Phare} (LP-t)}
\label{sec:methods_LePhare}
The public code \texttt{Le~Phare}
\citep{2002MNRAS.329..355A,2006A&A...457..841I} is primarily dedicated
to estimate photo-$z$'s, but it can also be used to estimate physical
parameters like stellar masses and infrared
luminosities. \texttt{Le~Phare} is based on a standard template
fitting procedure. The templates are redshifted and integrated through
the instrumental transmission curves. The opacity of the IGM is taken
into account and internal extinction could be added as a free
parameter to each galaxy. The photo-$z$'s are obtained by comparing
the modelled fluxes and the observed fluxes with a $\chi^2$ merit
function. A probability distribution function is associated to each
photo-$z$.

For the PHAT1 sample, we adopted a configuration similar to the one
used in the COSMOS field \citep{2009ApJ...690.1236I}:
\begin{itemize}
\item {\it Templates:} The set of templates was generated by
  \cite{2007ApJ...663...81P} with the code GRASIL
  \citep{1998ApJ...509..103S}.  The 9 galaxy templates of
  \cite{2007ApJ...663...81P} include 3 SEDs of elliptical galaxies and
  6 templates of spiral galaxies (S0, Sa, Sb, Sc, Sd, Sdm). Those were
  complemented with 12 additional blue templates generated with
  \cite{2003MNRAS.344.1000B}. Four different dust extinction laws were
  applied \citep[][and an additional bump at
    2175\AA]{1984A&A...132..389P,2000ApJ...533..682C}, depending on
  the considered template. Emission lines were added to the templates
  using relations between the UV continuum, the star formation rate
  and the emission line fluxes \citep{1998ARA&A..36..189K}.
\item {\it Prior:} No prior on the redshift distribution was
  applied. However, no redshift solution which would produce a galaxy
  brighter than $M(B)=-24$ was allowed. Such a prior would create
  catastrophic failure for some QSOs, but it was not explicitly
  intended to estimate photo-$z$'s for QSOs (no AGN templates were
  included in this run), although the PHAT1 catalogue contains some
  (see below).
\item {\it Training:} An automatic calibration of the zero-points was
  performed using the spec-z sample. The calibration is obtained by
  comparing the observed and modelled fluxes
  \citep{2006A&A...457..841I}. The calibration is done iteratively
  until convergence in the zero-points values is reached. This step
  helps in removing bias.
\end{itemize}

\subsection{\texttt{LRT} (LR-t)}
\texttt{LRT} \citep[Low-Resolution Spectral Templates
][]{2008ApJ...676..286A,2010ApJ...713..970A} is a set of subroutines
intended for estimating K-corrections and photometric redshifts using
a basis of empirical low resolution SED templates (hence \texttt{LRT})
for galaxies and AGNs. In this basis, every galaxy is represented by a
non-negative linear combination of three empirically determined SED
templates that resemble an elliptical, an Sbc spiral and an Im
irregular galaxy. Given the nature of the tests in the PHAT
initiative, the AGN SED template was not used. For the PHAT0 testing
phase, the \texttt{LRT} subroutines were modified to do a simple
$\chi^2$ minimisation to fit each template to the data separately
rather than fitting a non-negative combination of them.

\begin{itemize}
\item {\it Templates:} The templates were derived from the extensive
  broad-band and spectroscopic observations of the NOAO Deep
  Wide-Field Survey \citep{1999ASPC..191..111J} Bo\"otes field and
  range in wavelength between 0.03 and 30$\mu$m. In the PHAT1 testing
  phase, the \texttt{LRT} subroutines were used with the SED templates
  derived in \citet{2008ApJ...676..286A} which have a shorter
  wavelength range (0.1--10$\mu$m) than the newer versions presented
  in \citet{2010ApJ...713..970A}. These newer SED templates also
  integrate an AGN component with variable extinction.
\item {\it Prior:} For estimating photometric redshifts, the
  \texttt{LRT} subroutines also use a simple luminosity function
  prior, which is by default based on the $R$-band luminosity function
  of \citet{1996ApJ...464...60L}.
\item {\it Training:} No training with the model-$z$'s/spec-$z$'s was
  performed.
\end{itemize}

\subsection{Purger (Template Repair) (PT-t)}
Originated from the template-based method described in
\cite{2003AJ....125..580C}, this method uses synthetic colours
calculated from the given spectral energy distribution templates. A
common approach for template fitting is to take a small number of
spectral templates T and choose the best fit by optimising the
likelihood of the fit as a function of redshift, type, and luminosity,
p(z, T, L).  Here a variant of this method is used that incorporates a
continuous distribution of spectral templates, enabling the error
function in redshift and type to be well defined. 

\begin{itemize}
\item {\it Templates:} This code is only run on PHAT0 so that no
  individual template set is associated with this code.
\item {\it Prior:} No prior was used.
\item {\it Training:} No training with the model-$z$'s was performed.
\end{itemize}

\subsection{\texttt{ZEBRA} (ZE-t \& ZE2-t)}
\texttt{ZEBRA} \cite[Zurich Extragalactic Bayesian Redshift
  Analyzer][]{2006MNRAS.372..565F} is a freely available, open source
photometric redshift code based on a SED template-fitting
approach. Built on top of a traditional Maximum Likelihood ansatz it
introduces and combines several novel methods that help to improve the
accuracy of photometric redshift estimates for galaxies and AGNs
\citep[see e.g.][for some recent
  applications]{2010ApJ...714L..47O,2010ApJS..187..560L}. First,
\texttt{ZEBRA} is able to detect and correct photometric offsets in
the input catalogue.  Second, \texttt{ZEBRA} can use spectroscopic
redshifts on a small fraction of the photometric sample to iteratively
correct the original set of input templates. This template correction
step has been shown to be a crucial ingredient in decreasing the bias,
the scatter, and the number of outliers in the redshift estimation
\citep[e.g.][]{2006MNRAS.372..565F,2007ApJS..172..117M}. Third, when
run in Bayesian mode \texttt{ZEBRA} computes the prior in
redshift-template space in a self-consistent manner from the input
catalogues and the redshift-template likelihood functions.  This prior
is consequently used to derive the posterior probability distribution
of each input object.  Here, since \texttt{ZEBRA} participates only in
PHAT0, it is run in its basic Maximum Likelihood mode and with the
provided templates. The following set of parameters are used. The
redshifts are allowed to vary in steps of 0.002 from 0 to 4. The
filter bands are mildly smoothed using a top-hat filter with FWHM of
$20\AA$. Finally, the spectral flux densities weighted with photon
energy, not photon counts, are computed using the --flux-type=1
option. For the ZE2-t runs the redshift stepping is reduced to 0.001
and no smoothing of the filter bands is performed.

\begin{itemize}
\item {\it Templates:} \texttt{ZEBRA} is only run on PHAT0 so that no
  individual template set is associated with this code.
\item {\it Prior:} No prior was used.
\item {\it Training:} No training with the model-$z$'s was performed.
\end{itemize}

\subsection{\texttt{ANN$z$} (AN-e)}
\texttt{ANN$z$} \citep{2004PASP..116..345C} is an empirical photo-$z$
code based on artificial neural networks. Such a network is made up of
several layers, each consisting of a number of nodes. The first layer
receives the galaxy magnitudes as inputs, while the last layer outputs
the estimated photometric redshift. The layers in between could
consist of any number of nodes each. The nodes are inter-connected,
and every connection carries a 'weight', which is a free parameter in
the parametrisation. When a network is trained the weights of all node
connections are determined by minimising a cost function $E$. To avoid
an over-fitting, every network is tested on a validation set of
galaxies, whose spectroscopic redshifts are also known. The network
with lowest value of $E$ as calculated on the validation set is
selected and the photometric sample is run through it for redshift
estimation.  An error bar is assigned to each photo-$z$ via a chain
rule \citep[see][for details]{2004PASP..116..345C}.  Neural networks
have been used e.g. for estimation of photo-$z$'s for the SDSS
\citep{2007MNRAS.375...68C,2008ApJ...674..768O,2008arXiv0812.3831A},
as well as forecasts of photometric redshifts for future surveys like
the Dark Energy Survey \citep{2008MNRAS.386.1219B} and Euclid
\citep{2008MNRAS.387..969A}.

A neural network architecture of N:2N:2N:1 was used for the PHAT tests
where N is the number of filters for which there are input
magnitudes. Different architectures were tested, but this did not lead
to any substantial improvement in the results. The choice of
architecture is fully justified by tests done in
\cite{2003MNRAS.339.1195F} and \cite{2004PASP..116..345C}.

\subsection{\texttt{BDT} (DT-e)}
The Boosted Decision Tree (\texttt{BDT}) algorithm
\citep{2010ApJ...715..823G} is a training-set-based method that
combines an ensemble of weak classifiers into a single, powerful
classifier. The spectroscopic training set is first divided into
redshift bins whose width is approximately half the expected photo-$z$
resolution of the algorithm for the given sample. We have found that a
finer binning choice does not improve the resolution.  For each bin, a
set of trees is trained intended to recognise as ``signal" those
galaxies whose redshift falls within the bin in question, and
``background" those that fall more than 2$\sigma$ away from the signal
bin, where $\sigma$ is the iteratively-determined photo-$z$
resolution. As training variables we use the observed magnitudes in
each band. The process of constructing an individual tree begins with
a root node containing all the training galaxies. The root node is
then split into two subsamples by placing a cut on the one variable
that best separates the sample into signal and background. Each new
node is subsequently split in this way until the nodes reach a certain
minimum size. The result is a tree containing nodes with predominantly
signal and predominantly background galaxies. The process of
``boosting" iteratively repeats this process, giving higher weight to
galaxies that were initially misclassified. The overall signal
probability of a galaxy is then obtained by combining the
classification output from approximately 50 trees in each photo-$z$
bin, where higher weight is given trees with lower misclassification
rates in the training set.

The method produces a photo-$z$ probability for each galaxy as a
function of redshift. This method therefore yields not only an
estimate of the best photo-$z$ and error, but a reconstruction of the
full redshift PDF, $P(z)$. In \citet{2010ApJ...715..823G} it was shown
that the BDT algorithm improves upon the default photo-$z$'s in the
SDSS spectroscopic sample, and that the PDFs yield a more accurate
reconstruction of the redshift distribution $N(z)$.

\subsection{Wolf (Empirical $\chi^2$) (EC-e)}
The method of \cite{2009MNRAS.397..520W} derives PDFs from empirical
models and is a subclass of kernel regression methods. It mimics a
template-based $ \chi^2$-technique with the main difference that an
empirical dataset is used in place of the template grid. Each object
in the empirical set contributes to the observed object with a
quantified probability. The PDF of redshifts thus obtained can be used
in its entirety or investigated for ambiguities. Here, it is just
reduced to an expectation value and RMS in redshift. Any kernel
approach requires to choose a kernel function which also acts as a
smoothing scale to the discrete empirical model grid. Here, we used a
Gaussian kernel function with $\sigma_m = 0\fm 1$. However, a
$\chi^2$-method is correctly implemented if the kernel function
applied to the model makes its density distribution match that of the
observed sample \citep[see the matched error scale in Sect. 6 of][for
  details]{2009MNRAS.397..520W}. As a consequence, redshift
distributions of object samples can be reconstructed potentially
accurate within Poisson noise of the sample sizes, which would also
imply no bias exceeding random noise.

\subsection{Purger (Nearest-Neighbour Fit) (PN-e)}
This empirical method compares the observed colours to the reference
set. The estimation method first searches the colour space for the $k$
nearest neighbours of every object in the estimation set (i.e. the
galaxies for which we want to estimate redshift) and then estimates
the redshift by fitting a local low order polynomial to these
points. An improved version of this code is using a k-d tree index for
fast nearest neighbour search \citep{2007AN....328..852C}. It was used
to calculate photometric redshifts for the SDSS Data Release 7
\citep{2009ApJS..182..543A}. The advantage of this method versus a
template-based method might be the better estimation accuracy, but it
cannot extrapolate, so the completeness of the reference set is
crucial. For this reason, we have used the large training set
available for the PHAT0 test.

The estimation was done using the large, simulated data set using 150
nearest neighbours. A small number of outliers was automatically
excluded from the regression on the neighbour sets.

\subsection{Li (Polynomial) (PO-e)}
\label{sec:code_PO}
This empirical photo-$z$ method is based on
\cite{2008AJ....135..809L}, which uses a polynomial fit so that the
galaxy redshift is expressed as the sum of its magnitudes and colours.
Different from \cite{2008AJ....135..809L} where the training set
galaxies are divided into several fixed colour-magnitude cells, here
the coefficients of the photo-$z$ polynomial are derived individually
for each galaxy by choosing a subset of training set galaxies whose
magnitudes and colours are closest to the input galaxy.  They are
chosen based on quadratically summed ranks of colour and magnitude
differences between the training set galaxies and the input
galaxy. All magnitudes and independent colours are used.  Note that
each training set galaxy has an equal weight in the fit. This may
introduce a redshift bias to input galaxies near the edges of the
colour-magnitude distributions.  Therefore, a better approach would be
to assign weights to the chosen training-set galaxies based on the
inverse value of their final rank, but this has not been implemented
for PHAT.

\subsection{Carliles (Regression Trees) (RT-e)}
The RT-e method by \cite{2010ApJ...712..511C} is based on Random
Forests which are an empirical, non-parametric regression technique.
A Random Forest builds an ensemble average of randomised regression
tree redshift estimates.  Bootstrap samples are created by sampling
from the training set with replacement, and each regression tree is
trained on its own bootstrap sample.  Given a new test object, each
regression tree produces its own redshift estimate, and these
estimates are averaged to yield the final Random Forest redshift
estimate.  This technique also results in Gaussian errors, and this
behaviour has a strong theoretical statistical explanation.
Intuitively speaking, a given new galaxy can be considered to be drawn
from the space of inputs (colours, magnitudes, etc.) by redshifts.
This space is the event space, and for that new galaxy one can
hypothesise the existence of a distribution over the event space,
unique to that galaxy, which reflects the similarity of the new galaxy
(minus the unknown redshift) to any given point in the event space.
The Random Forest approximates this distribution per object, and the
process results in easily computable per-object error parameter
estimates.

For the PHAT tests a leaf size of 5 was chosen and 50 trees were used.

\subsection{Singal (Neural Network) (SN-e)}
The primary motivation for the development of this code was to treat
additional available galaxy information beyond photometric data, for
example shape parameters, on an equal footing with the photometric
data \citep[as it was done in e.g. ][]{2004PASP..116..345C,
  2004MNRAS.348.1038B}. The package, although still undergoing
modification, is a multi-layer perceptron neural network for the IDL
environment.  The IDL code can be relatively easily modified, and
could in principle be optimised for a variety of input data
situations.  As training convergence is relatively slow in this
network, it is most useful in situations where a robust training set
is available from the outset.

As implemented here, the network has an input layer of neurons which
accepts the magnitudes in each band.  The input layer treats all input
information on an equal footing, normalising across all objects in the
training set so that the inputs for each neuron on the input layer are
distributed between 0 and 1. There are two hidden layers of 30 neurons
each, and an output layer with a single neuron obtaining a value
between 0 and 1 which is a proxy for the estimated redshift, with the
linear conversion defined during the training when the known redshifts
of the training set are supplied subject to the conversion.

\begin{table*}
\begin{minipage}[t]{\textwidth}
\caption{Methods used for photo-$z$ estimation within PHAT}
\label{tab:methods}
\centering
\renewcommand{\footnoterule}{}  % to avoid a line before footnotes
\begin{tabular}{llllc}
\hline
\hline
Acronym & Participant & Code & Reference & Public\\
\hline
BP-t & Coe, D.         & \texttt{BPZ}, Bayesian Photometric Redshifts & \cite{2000ApJ...536..571B,2006AJ....132..926C} & $\surd$\footnote{\url{http://acs.pha.jhu.edu/~txitxo/} ; version 1.99.3 used for PHAT: \url{http://www.its.caltech.edu/~coe/BPZ/}}\\
BP2-t& Benitez, N.     & \texttt{BPZ}, Bayesian Photometric Redshifts & \cite{2000ApJ...536..571B}; Ben\'itez 2010 in prep. & $\surd^a$\\
EA-t & Brammer, G.     & \texttt{EAZY}, Easy and Accurate Redshifts from Yale & \cite{2008ApJ...686.1503B} & $\surd$\footnote{\url{http://www.astro.yale.edu/eazy/}}\\
GA-t & Kotulla, R.     & \texttt{GALEV}, GALaxy EVolution & \cite{2009MNRAS.396..462K} & $\surd$\footnote{\url{http://www.galev.org/}}\\
GO-t & Dahlen, T.      & \texttt{GOODZ} & \cite{2005ApJ...631..126D,2007ApJ...654..172D} & \\
HY-t & Miralles, J.-M. & \texttt{Hyperz} & \protect{\cite{2000A&A...363..476B}} & $\surd$\footnote{\url{http://webast.ast.obs-mip.fr/hyperz/}}\\
KR-t & Schmidt, S.     & \texttt{Kernelz}, Kernel Regression & Schmidt \& Brewer (in prep) & \\
LP-t & Arnouts, S.     & \texttt{Le~Phare} & \protect{\cite{2006A&A...457..841I}} & $\surd$\footnote{\url{http://www.cfht.hawaii.edu/~arnouts/lephare.html}}\\
     & Ilbert, O.      & & & \\
LR-t & Assef, R.       & \texttt{LRT}, Low-Resolution Spectral Templates & \cite{2008ApJ...676..286A,2010ApJ...713..970A} & $\surd$\footnote{\url{http://www.astronomy.ohio-state.edu/~rjassef/lrt/}}\\
PT-t & Purger, N.      & Template Repair & \cite{2007ApJS..172..634A} & $\surd$\footnote{\url{http://skyserver.elte.hu/PhotoZ/}} \\
ZE-t & Feldmann, R.    & \texttt{ZEBRA}, Zurich Extragalactic Bayesian Redshift Analyzer & \cite{2006MNRAS.372..565F} & $\surd$\footnote{\url{http://www.exp-astro.phys.ethz.ch/ZEBRA/}}\\
ZE2-t& Gillis, B.      & \texttt{ZEBRA}, Zurich Extragalactic Bayesian Redshift Analyzer & \cite{2006MNRAS.372..565F} & $\surd^h$\\
\hline
AN-e & Abdalla, F.     & \texttt{ANN$z$}, Artificial Neural Network & \cite{2004PASP..116..345C} & $\surd$\footnote{\url{http://www.homepages.ucl.ac.uk/~ucapola/annz.html}}\\
     & Banerji, M.     & & & \\
DT-e & Gerdes, D.      & \texttt{BDT}, Boosted Decision Trees & \cite{2010ApJ...715..823G} & \\
EC-e & Wolf, C.        & Empirical $\chi^2$ & \cite{2009MNRAS.397..520W} & \\
PN-e & Purger, N.      & Nearest-Neighbour Fit & \cite{2009ApJS..182..543A} & $\surd^{\rm g}$\\
PO-e & Li, I.~H.       & Polynomial Fit & \cite{2008AJ....135..809L} & \\
RT-e & Carliles, S.    & Regression Trees & \cite{2010ApJ...712..511C} & $\surd$\footnote{\url{http://www.sdss.jhu.edu/~carliles/photoZ/}}\\
SN-e & Singal, J.      & Neural Network & - & $\surd$\footnote{\url{http://www.slac.stanford.edu/~jacks/}}\\
\hline
\end{tabular}
\end{minipage}
\end{table*}

\section{PHAT0 - a highly idealised simulation}
\label{sec:PHAT0}

\subsection{Motivation}
The lowest algorithmic level of the codes can be tested if the
photometry is bias-free and everything except for the redshifts is
provided. In this way the choice of template sets, the use of priors,
etc. do not play a role and code-specific problems can be disentangled
from other effects. To this end, simulations with synthetic photometry
are set up with the LP-t photo-$z$ code (see
Sect.~\ref{sec:methods_LePhare}).

\subsection{Data set}
\label{sec:PHAT0_data}
In order to keep things simple PHAT0 is based on a very limited
template set and a long wavelength baseline. A noise-free catalogue
with accurate synthetic colours is provided as well as a catalogue
with a low level of additional noise. Furthermore, we added a very
large training set to ensure that also empirical photo-$z$ algorithms
find an ideal environment. The ingredients are detailed in the
following. 

Everything but the redshifts for the test data set was revealed to the
participants. In particular, the template set
(Sect.~\ref{sec:PHAT0_template_set}) and the filter curves
(Sect.~\ref{sec:PHAT0_filter_set}) were provided, and details about
the construction of the catalogues
(Sect.~\ref{sec:noise_free_cat}~\&~\ref{sec:noise_cat}; e.g. the used
IGM recipe) were revealed. The participants were explicitly asked to
use those ingredients if applicable to make their setup as comparable
to the simulation setup as possible.

\subsubsection{Template set}
\label{sec:PHAT0_template_set}
\begin{figure}
\resizebox{\hsize}{!}{\includegraphics[width=\textwidth]{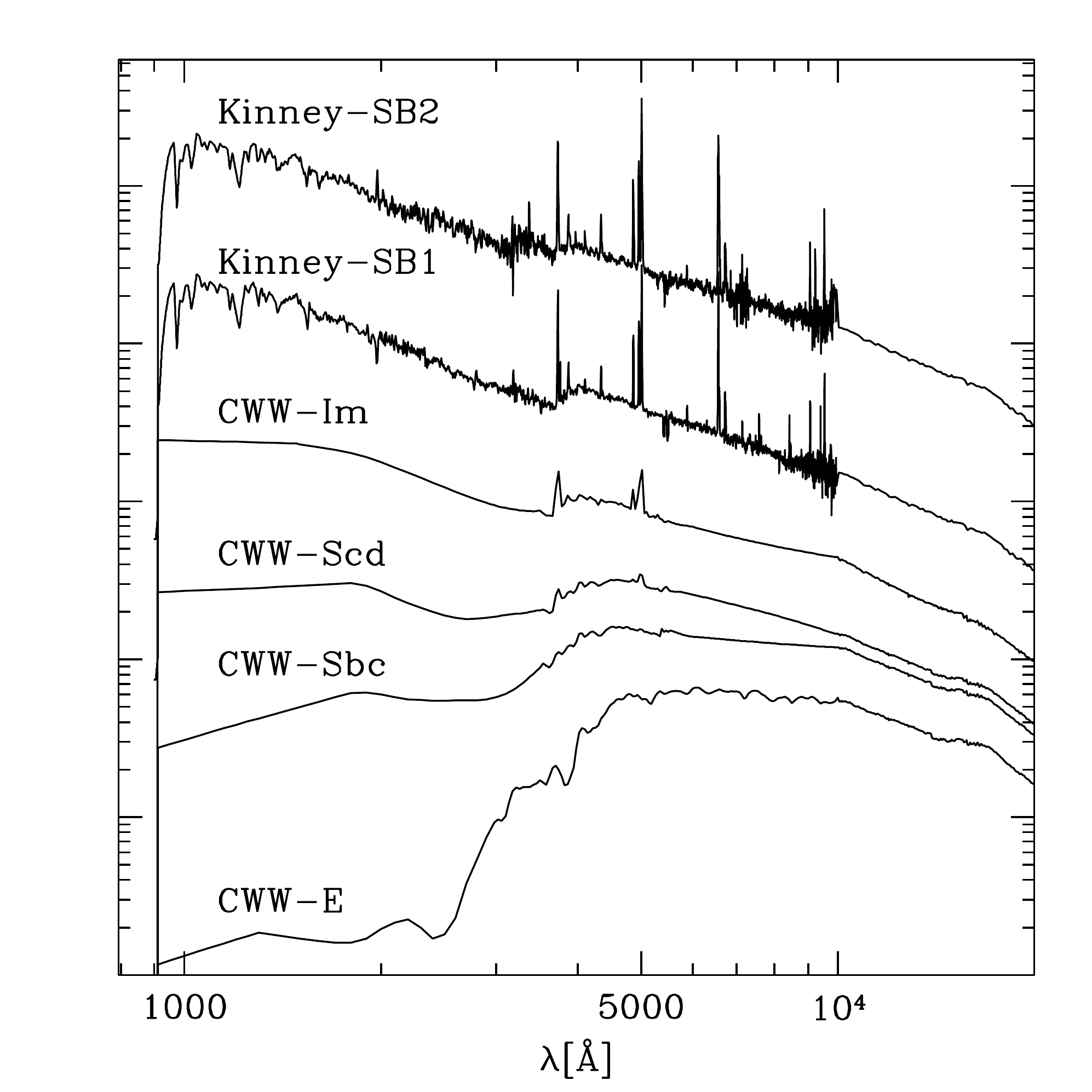}}
\caption{Template set used for the PHAT0 test (arbitrary flux
  normalisation).}
\label{fig:templates}
\end{figure}

The empirical template set by \cite{1980ApJS...43..393C} has been used
extensively in different photo-$z$ studies. As in the case of
LP-t \citep{2006A&A...457..841I} and BP-t
\citep{2000ApJ...536..571B} we decided to supplement this template set
by two templates for starburst galaxies from
\cite{1996ApJ...467...38K}. The template SEDs are displayed in
Fig.~\ref{fig:templates}.

It should be noted that the choice of the template set is not critical
in this test because the template set is provided to the participants
using template-based codes and the very large training set (see below)
covers densely the whole SED-redshift space. This particular set is
chosen here because it is one of the most widely used sets for
photo-$z$'s in its original, extended, and modified (re-calibrated)
form. Participants using template-based codes were explicitly asked to
use this particular template set for the PHAT0 test and switch off any
priors within their codes.

\subsubsection{Filter set}
\label{sec:PHAT0_filter_set}
For the PHAT0 test we want to avoid systematic effects that can arise
in photo-$z$'s because of an insufficient coverage in wavelength. For
example, colour-redshift degeneracies \citep[see
  e.g. ][]{2000ApJ...536..571B} can occur between high- and
low-redshift if infrared (IR) and/or ultraviolet (UV) bands are not
available.

\begin{figure}
\resizebox{\hsize}{!}{\includegraphics[width=\textwidth]{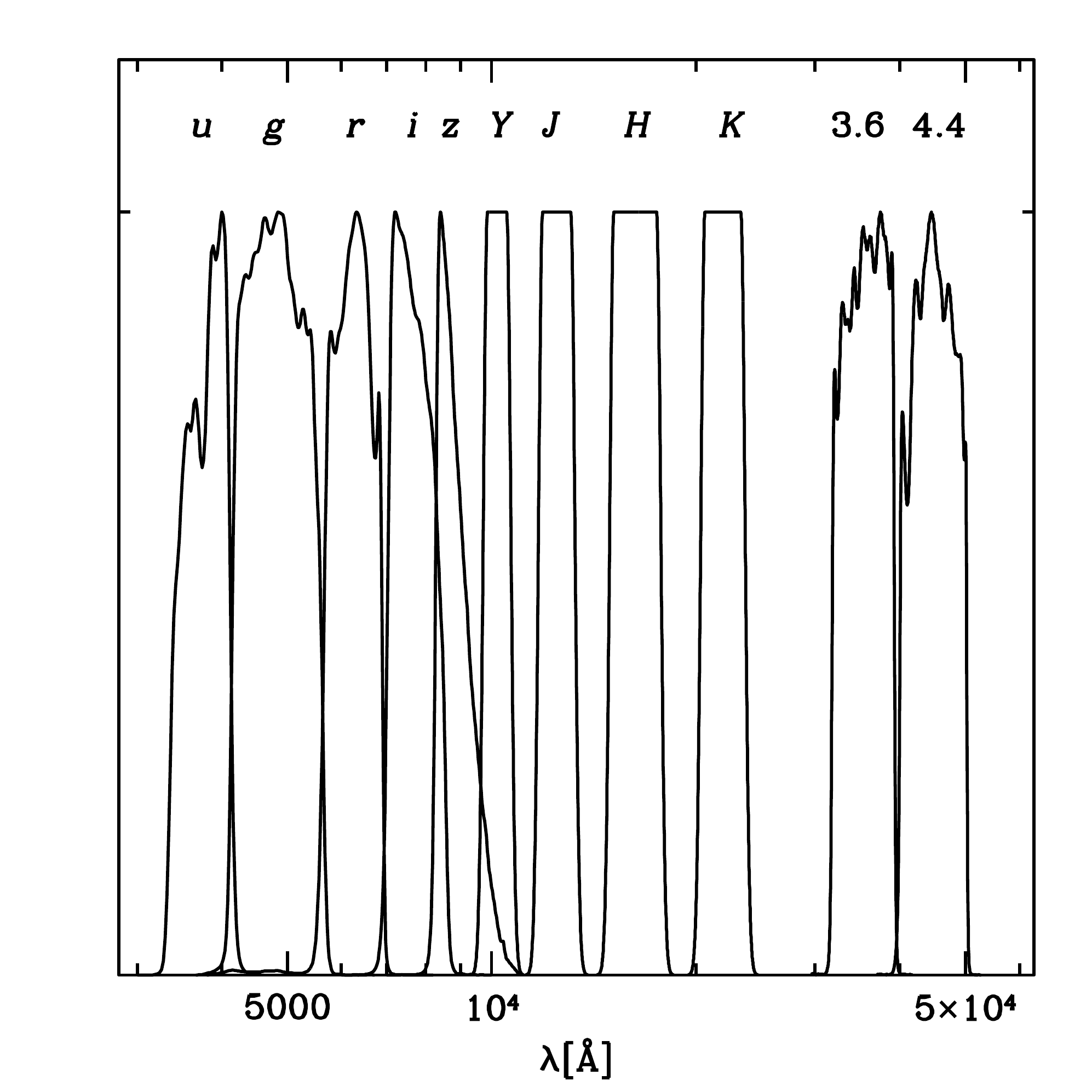}}
\caption{Transmission curves of the filter set used for the PHAT0 test}
\label{fig:filters}
\end{figure}

Thus, the filter set used here spans the whole range from near-UV to
mid-IR (see Fig.~\ref{fig:filters}). We choose the $ugriz$-bands from
MEGACAM mounted at the CFHT \citep{2003SPIE.4841...72B}, the
$YJHK$-bands of UKIDSS \citep{2007MNRAS.379.1599L}, and the two bluer
bands of the IRAC camera mounted on the Spitzer Space Telescope
\citep{2004ApJS..154...10F}. Again this choice is not too critical
since the filter curves are provided and one of the tests does not
include any noise at all and the other one includes just a low level
of noise in the photometry.

\subsubsection{Noise-free catalogue}
\label{sec:noise_free_cat}
One of the most simple tests one can think of is to compare the
redshift estimates of different codes for data with infinite
signal-to-noise (S/N) and thus perfect colours. In this way the
agreement of the basic interpolation- and convolution-algorithms in
template-based codes can be tested. Any differences found in such a
basic test will probably propagate to more realistic setups.

We use the LP-t code as a reference to create such a catalogue evenly
distributed over the six templates and over the redshift range $0<z<4$
including the effect of absorption by the intergalactic medium (IGM)
following the recipe by \cite{1995ApJ...441...18M}. The model
redshifts were revealed to the participants for this test.

It should be noted that inaccurate redshift estimates from one of the
codes only mean that this particular code does not agree perfectly
with LP-t. Which of the two codes is inaccurate (or
whether even both are inaccurate) cannot be decided with such a test.

\subsubsection{Catalogue with noise}
\label{sec:noise_cat}
\begin{figure}
\resizebox{\hsize}{!}{\includegraphics[width=\textwidth]{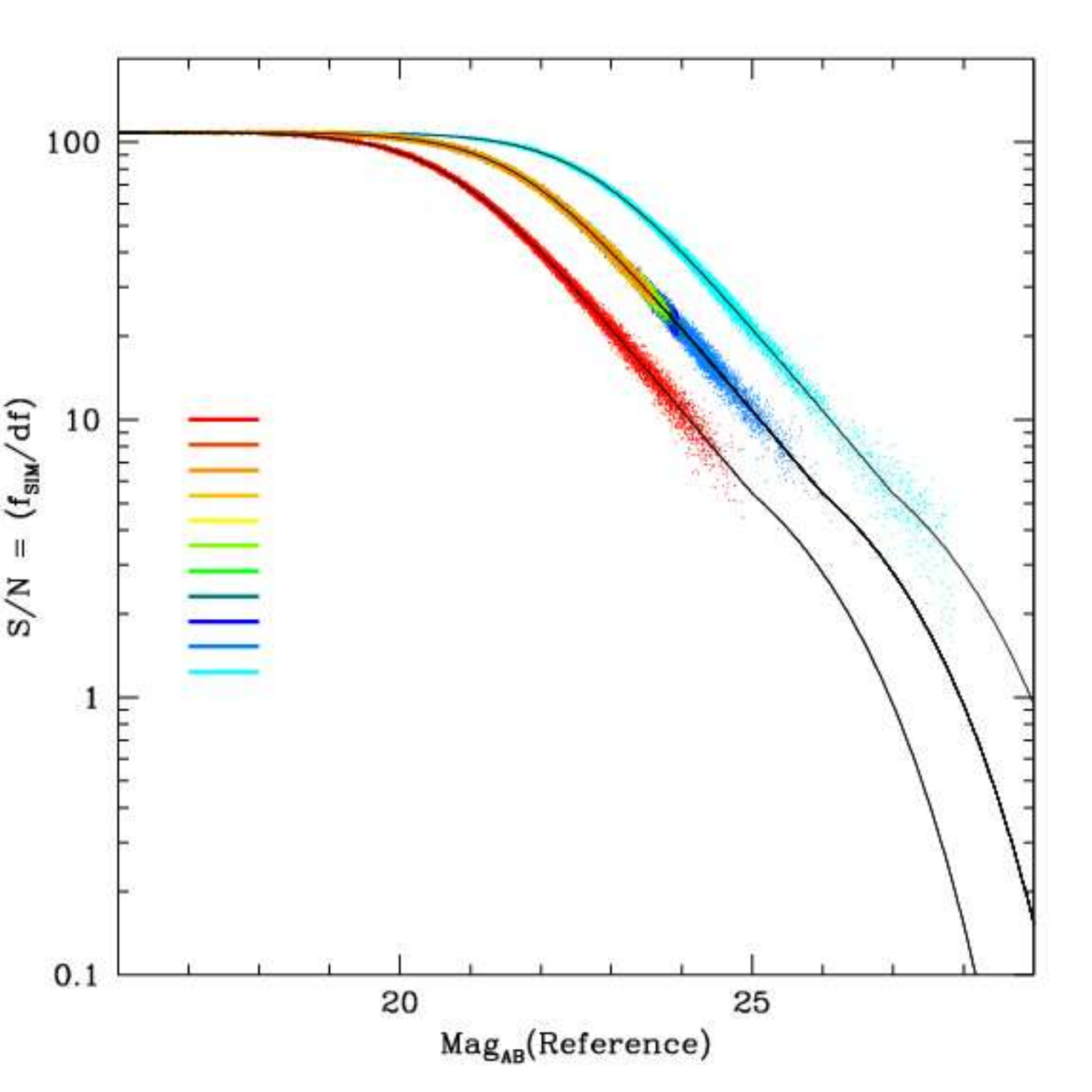}}
\caption{Signal-to-noise model used for the PHAT0 test}
\label{fig:S_N}
\end{figure}

To study the influence of noise on the results, a more realistic
catalogue is set up as well. We adopt a parametric form for the
signal-to-noise as a function of magnitude which behaves as a
power-law at bright magnitudes and an exponential at faint
magnitudes. The transition regime is defined by the parameters
($m_{\star}, err_{\star}$).  At magnitude $m\le m_{\star}$, we adopt
$err(m)=10^{0.4(\alpha_{bright}+1)(m-m_{\star})}$, and at magnitude
$m\ge m_{\star}$, we use
$err(m)=\frac{err_{\star}}{2.72}.exp(10^{\alpha_{faint}(m-m_{\star})})$,
where $\alpha_{bright}$ and $\alpha_{faint}$ are the slopes at bright
and faint magnitudes respectively.  The adopted values for each filter
are reported in Table~\ref{tab:filters}, while the behaviour of the
Signal-to-Noise (${\rm S/N}=1.086/err$) for the different passbands is
shown in Fig~\ref{fig:S_N} (colour coded from u band, in cyan to
4.5$\mu m$, in red).  The noisy magnitudes are randomly drawn assuming
a Gaussian distribution in flux with mean and standard deviation
($flux,err(flux)$).

To generate the simulated catalogue, the galaxies are distributed
according to $r$-band luminosity functions for the different spectral
types. However, for simplicity in the comparison of the different
codes, we do not apply any dust attenuation for the star-forming
galaxies and we do not let the luminosity functions evolve with
redshift. Thus, this simulated catalogue is not expected to provide a
realistic distribution of low and high redshift galaxies.  Note, that
we do include the averaged Lyman absorption by the intergalactic medium
as a function of redshift, following \cite{1995ApJ...441...18M} which
will affect the blue bands at high redshift. The catalogue has been
cut to objects brighter than $r=24$, so that only reasonably high-S/N
sources are included. The redshift distribution attains a smooth shape
with a peak at intermediate redshifts and few objects beyond $z=1.5$.

The final catalogue consists of $\sim11\,000$ objects for which the
redshifts are not revealed to the participants. Furthermore, a much
larger training set of $\sim170\,000$ objects with exactly the same
properties as the original catalogue is provided.

\begin{table}
\caption{Filters used for the PHAT0 test}
\label{tab:filters}
\centering
\begin{tabular}{llrrrr}
\hline
\hline
Filter & Instrument & $m_\star$ & $\rm err_\star$ & $\alpha_{\rm bright}$ & $\alpha_{\rm faint}$ \\
\hline
$u$        & MEGACAM@CFHT   & 27.0 & 0.2 & $-0.25$ & 0.22\\
$g$        & MEGACAM@CFHT   & 26.0 & 0.2 & $-0.25$ & 0.22\\
$r$        & MEGACAM@CFHT   & 26.0 & 0.2 & $-0.25$ & 0.22\\
$i$        & MEGACAM@CFHT   & 26.0 & 0.2 & $-0.25$ & 0.22\\
$z$        & MEGACAM@CFHT   & 26.0 & 0.2 & $-0.25$ & 0.22\\
$Y$        & WFCAM@UKIRT    & 26.0 & 0.2 & $-0.25$ & 0.22\\
$J$        & WFCAM@UKIRT    & 26.0 & 0.2 & $-0.25$ & 0.22\\
$H$        & WFCAM@UKIRT    & 26.0 & 0.2 & $-0.25$ & 0.22\\
$K$        & WFCAM@UKIRT    & 26.0 & 0.2 & $-0.25$ & 0.22\\
$3.6\mu m$ & IRAC@Spitzer   & 25.0 & 0.2 & $-0.25$ & 0.22\\
$4.5\mu m$ & IRAC@Spitzer   & 25.0 & 0.2 & $-0.25$ & 0.22\\
\hline
\end{tabular}
\end{table}

\subsection{Results for the noise-free case}
In the following we will present the results of three different
template-based codes on the noise-free catalogue that were submitted
after the release. The training of empirical codes on noise-free data
often does not make sense. That is probably the reason why no results
for empirical codes on the noise-free data have been submitted to
PHAT.

\begin{figure*}
\caption{Results of the PHAT0 test for the noise-free catalogue}
\centering
\includegraphics[width=17cm]{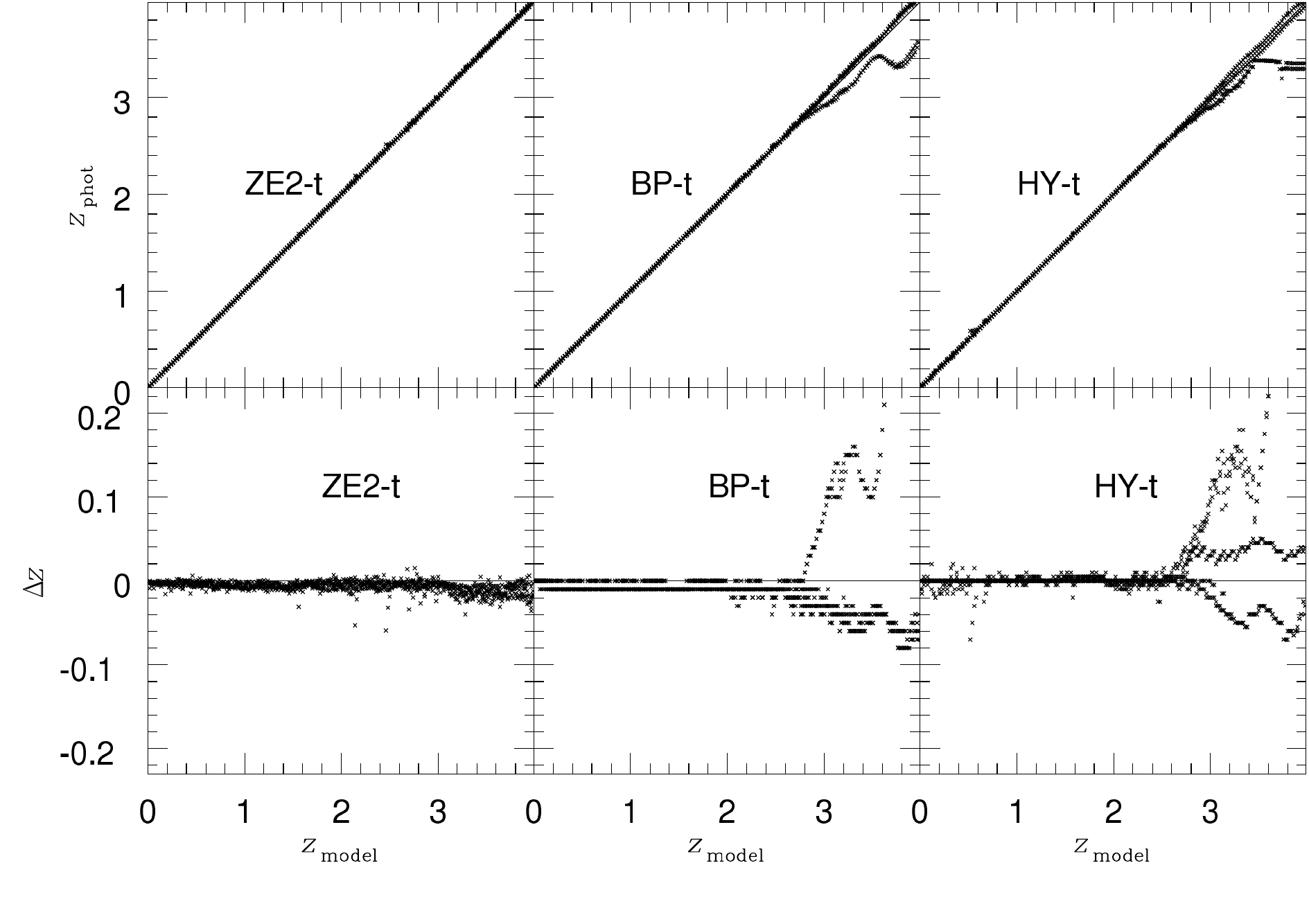}
\label{fig:res_noisefree}
\end{figure*}

The results are summarised in Fig.~\ref{fig:res_noisefree} showing the
model redshift $z_{\rm model}$ against the redshift estimate $z_{\rm
  phot}$ and the redshift difference $\Delta z = z_{\rm model} -
z_{\rm phot}$.

The ZE2-t code shows nearly perfect agreement with LP-t in this test
in terms of redshift estimates. This suggests strongly that the basic
interpolation of the filter- and template-curves and their subsequent
convolution by the two codes leads to colour estimates that agree very
well. Also the modelled attenuation of the IGM seems to be identical
in both codes.

Up to a redshift of $z\sim2.5$ the agreement between LP-t
and HY-t/BP-t is close to perfect as well. For
higher redshifts there are considerable discrepancies between
LP-t on the one hand and HY-t and BP-t
on the other hand.

\begin{figure}
\resizebox{\hsize}{!}{\includegraphics[width=\textwidth]{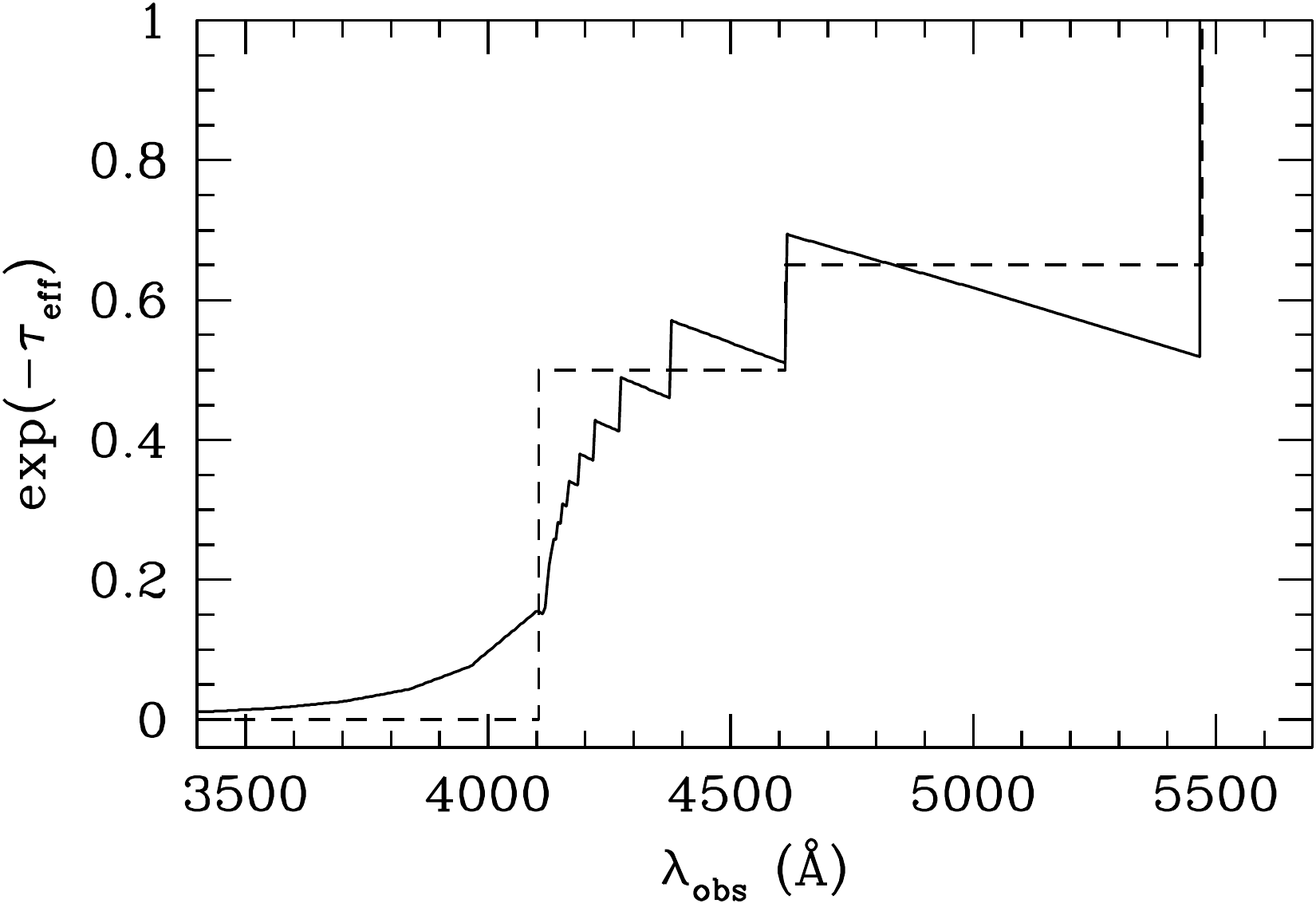}}
\caption{Opacity curves used by LP-t (solid) and
  HY-t and BP-t (dashed) for a redshift of
  $z=3.5$.}  \centering
\label{fig:opacity}
\end{figure}

A further analysis shows that especially the blue templates with
considerable UV flux get assigned grossly wrong redshift estimates. At
a redshift of $z\sim2.5$ the Lyman-$\alpha$ line enters our filter
set. These two facts suggest that the handling of the IGM, i.e. the
opacity of the Lyman-$\alpha$ forest, is implemented differently in
the codes. Although all codes refer to the paper of
\cite{1995ApJ...441...18M}, it turns out that HY-t and BP-t use an
analytic approximation of the opacity curve. As described in that
paper the opacity curve can be approximated by a step-function with
depression factors $D_A$ and $D_B$ shortward of Lyman-$\alpha$ and
Lyman-$\beta$, respectively, and a complete absorption shortward of
the Lyman-limit. LP-t uses the full opacity curve instead (binned for
redshift intervals of $\Delta z=0.1$). See Fig.~\ref{fig:opacity} for
a comparison of the opacity curves for a redshift of $z=3.5$.

The scatter around the mean opacity curve for a given redshift is
rather large \citep[see Fig.~3 of ][]{1995ApJ...441...18M} due to
clustering of the IGM. Thus, for practical applications we do not
expect either method to perform superior over the other one as long as
a direct relation between opacity and redshift is assumed. To account
for the greatly varying optical depth of the IGM for different
lines-of-sight at a fixed redshift in a realistic application, one
certainly would have to vary opacity as another free parameter. The
discrepancies reported here just appear in this artificial test
without noise and a fixed opacity-redshift relation. However,
different residuals between model and observation might well be
present in applications of photo-$z$ codes with a fixed
opacity-redshift relation to real data.

\subsection{Results for the catalogue with noise}
We select the best fit or most likely photo-$z$ estimate from each
method. Some methods provide estimates of confidence in their
photo-$z$'s in the form of redshift uncertainties or probability
distributions $P(z)$ and/or template quality of fit measurements like
$\chi^2$.  These can help identify and prune those photo-$z$ estimates
most likely to be outliers.  However these confidence measures are not
performed consistently or universally among the various methods, so we
do not consider them here.

\begin{figure*}
\caption{Results of the PHAT0 test for the catalogue with noise,  $z_{\rm phot}$ vs. $z_{\rm model}$. Note that LP-t (top-left panel) was used to create the simulations and should be regarded as a reference.}
\centering
\includegraphics[width=17cm]{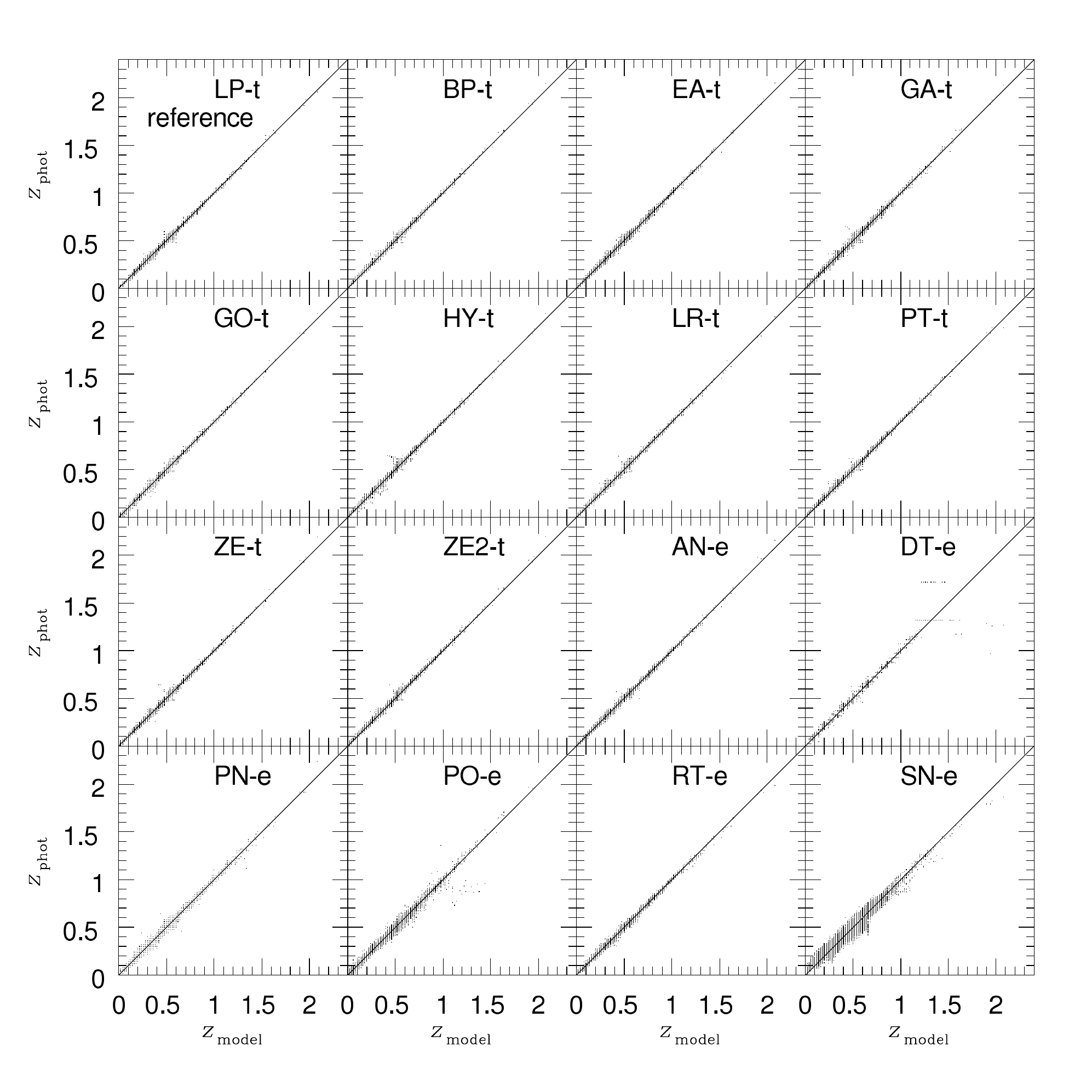}
\label{fig:res_noise}
\end{figure*}

\begin{figure*}
\caption{Results of the PHAT0 test for the catalogue with noise,
  $\Delta z=z_{\rm model}-z_{\rm phot}$ vs. $z_{\rm model}$. Note that
  LP-t (top-left panel) was used to create the simulations and should
  be regarded as a reference.}  \centering
\includegraphics[width=17cm]{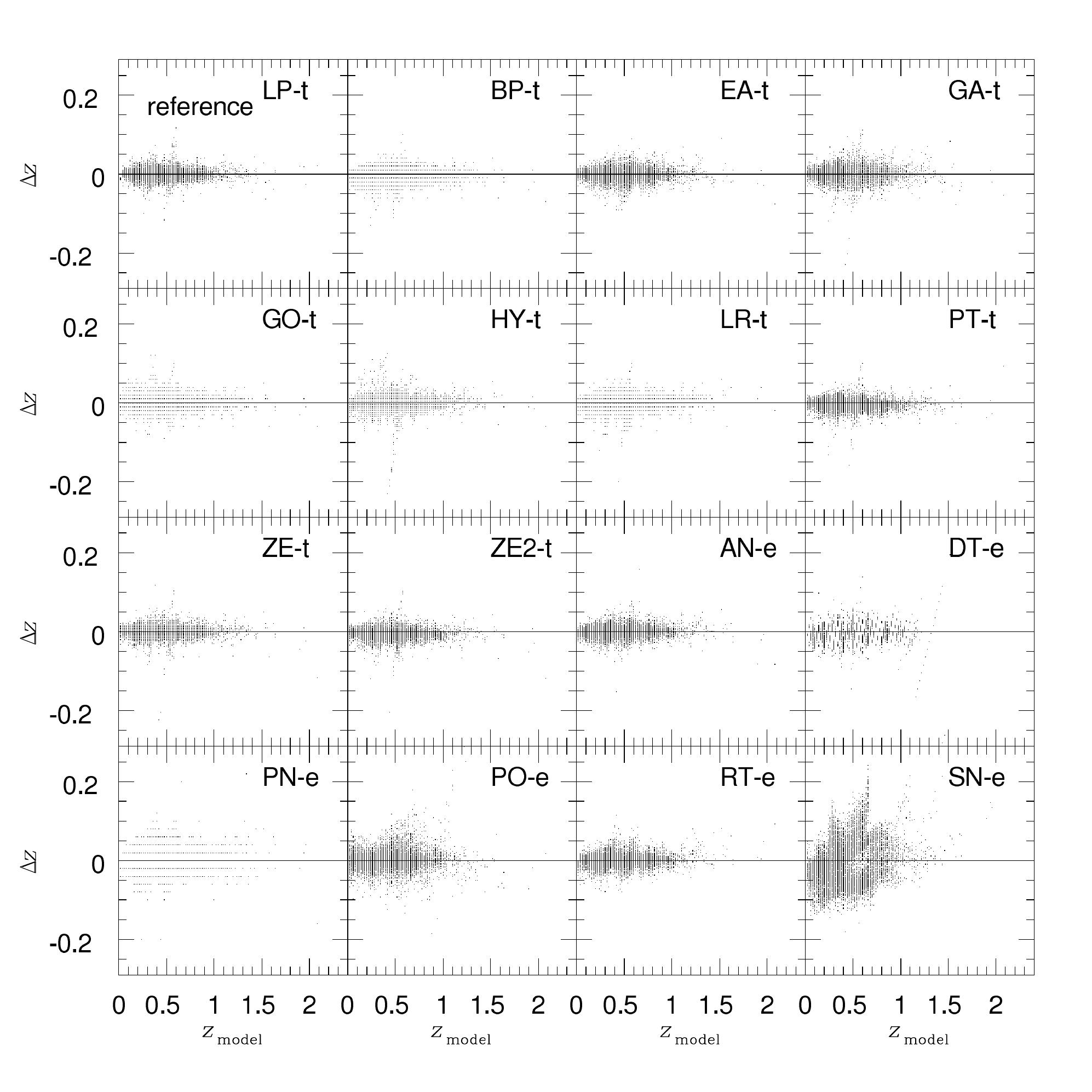}
\label{fig:res_noise_diff}
\end{figure*}

\begin{figure}
\resizebox{\hsize}{!}{\includegraphics[width=\textwidth]{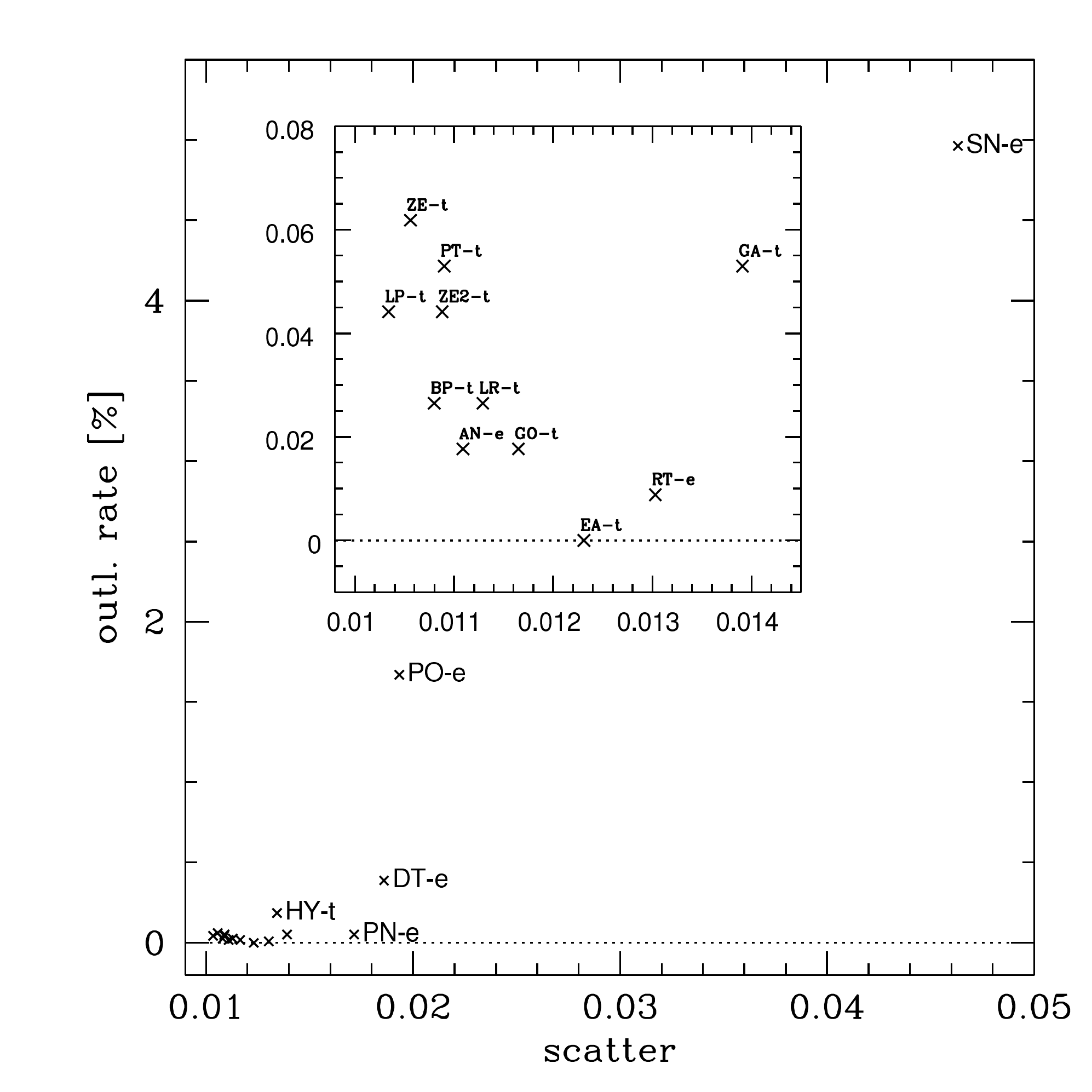}}
\caption{Scatter and outlier values for the catalogue with noise of
  PHAT0. The inlet shows the region in the lower left as a blow-up,
  but due to shot noise the performance of most the codes in the inlet
  should be regarded as identical.}
\label{fig:stats0}
\end{figure}

The error distribution of photo-$z$'s is usually non-Gaussian with
extended tails and some catastrophic outliers with grossly wrong
redshift estimates. To summarise this distribution by a few numbers is
not always possible. Here we express the photo-$z$ accuracy in terms
of the mean and the RMS scatter of the quantity $\Delta z = z_{\rm
  model}-z_{\rm phot}$ (after rejection of outliers), and an outlier
rate, as it was done in many former studies. These statistics for the
different codes can be found in
Table~\ref{tab:PHAT0_noise_res}. Figure~\ref{fig:stats0} shows the
scatter and outlier values in comparison. We define all objects with a
redshift estimate that differs by more than 0.1 from the model
redshift, i.e. $|\Delta z|=|z_{\rm model}-z_{\rm phot}|>0.1$, as
outliers. We refer the reader to the diagrams in
Figs.~\ref{fig:res_noise}~\&~\ref{fig:res_noise_diff} showing the
complete error distribution.

\begin{table}
\begin{minipage}[t]{\columnwidth}
\caption{Results for the PHAT0 catalogue with noise}
\label{tab:PHAT0_noise_res}
\centering
\renewcommand{\footnoterule}{}  % to avoid a line before footnotes
\begin{tabular}{lrrr}
\hline
\hline
Acronym & bias & scatter & outlier rate\footnote{Outliers are defined as objects with $|\Delta z|=|z_{\rm model}-z_{\rm phot}|>0.1$.} \\
\hline
LP-t & $ 0.000$ & $0.010$ & $ 0.044\%$ \\
\hline
BP-t & $-0.005$ & $0.011$ & $ 0.026\%$ \\
EA-t & $-0.001$ & $0.012$ & $ 0.000\%$ \\
GA-t & $ 0.000$ & $0.014$ & $ 0.053\%$ \\
GO-t & $ 0.000$ & $0.012$ & $ 0.018\%$ \\
HY-t & $-0.002$ & $0.013$ & $ 0.185\%$ \\
LR-t & $ 0.000$ & $0.011$ & $ 0.026\%$ \\
PT-t & $-0.005$ & $0.011$ & $ 0.053\%$ \\
ZE-t & $ 0.000$ & $0.011$ & $ 0.062\%$ \\
ZE2-t& $-0.005$ & $0.011$ & $ 0.044\%$ \\
\hline
AN-e & $ 0.000$ & $0.011$ & $ 0.018\%$ \\
DT-e & $-0.004$ & $0.019$ & $ 0.389\%$ \\
PN-e & $ 0.000$ & $0.017$ & $ 0.053\%$ \\
PO-e & $ 0.001$ & $0.019$ & $ 1.669\%$ \\
RT-e & $ 0.000$ & $0.013$ & $ 0.010\%$ \\
SN-e & $-0.005$ & $0.049$ & $18.202\%$ \\
\hline
\end{tabular}
\end{minipage}
\end{table}

\subsubsection{Results from LP-t (Arnouts)}
In order to set a standard to which the performance of all other codes
can be compared to, we run LP-t on the catalogue with
noise that was created by the code itself. It is reasonable to regard
the accuracy reached by LP-t on this catalogue as a
theoretical limit set by the amount of noise put in (see
Sect.~\ref{sec:noise_cat}). The results are displayed in the first
panels of Figs.~\ref{fig:res_noise}~\&~\ref{fig:res_noise_diff}
alongside the results from the other codes.

\subsubsection{Results from the other codes}
The numbers in Table~\ref{tab:PHAT0_noise_res} and the observed error
distributions displayed in
Figs.~\ref{fig:res_noise}~\&~\ref{fig:res_noise_diff} suggest that
most codes tested here perform similarly to LP-t. Note
that there is some degeneracy between the scatter values and the
outlier rates. No significant bias is produced by any of the
codes. All bias values are smaller than 0.5\%. Looking at the scatter
values and outlier rates four different groups can be identified:

\begin{enumerate}
\item A large number of codes (AN-e, BP-t, GO-t, EA-t, LR-t, RT-e,
  PT-t, ZE-t, ZE2-t) performs very similarly to LP-t with scatter
  values only slightly larger and outlier rates that are very similar
  or even smaller. This can be regarded as essentially identical
  performance because the low numbers of outliers are strongly
  affected by shot-noise. Note that the outlier rates of these codes
  correspond to $0-7$ out of $\sim11\,000$ objects!
\item Some other codes (GA-t, HY-t, PN-e) show larger values in both
  statistics than LP-t, but the differences are still
  minor and not very significant.
\item The codes DT-e, PO-e yield scatter values that are larger by a
  factor of two and outlier rates that are much larger than the
  LP-t statistics, with DT-e yielding a smaller outlier
  rate than PO-e.
\item SN-e performs worse but is still in the development phase.
\end{enumerate}
In the following we discuss the problems occurring in the last two
groups.

\subsubsection{Problems}
The panels for DT-e of
Figs.~\ref{fig:res_noise}~\&~\ref{fig:res_noise_diff} clearly show
that the code performs very similar to the codes from groups 1. \&
2. for redshifts $z_{\rm model}\la1.1$. For larger redshifts the
training set becomes more and more sparse. The division into branches
of the decision tree hence becomes less precise. For the highest
redshift interval only one branch is established so that objects from
a rather large range in $z_{\rm model}$ are all assigned the same
$z_{\rm phot}$. This particular feature of the DT-e code leads to the
slightly worse statistics reported in Table~\ref{tab:PHAT0_noise_res}.

The empirical code PO-e (see Sect.~\ref{sec:code_PO}) is based on a
second-order polynomial fit of the colour-redshift relation. This
leads to a very limited number of degrees of freedom (66 in the PHAT0
case with 11 bands) compared to the number of objects in the training
set.\footnote{Note that PO-e was trained on a much smaller training
  set with $\sim1200$ objects.} Not all the information included in
the training set can be reflected by the 66 coefficients so that this
empirical code performs worse in this test than other empirical codes
(e.g. AN-e) that feature many more degrees of freedom.

The SN-e code was developed for a low redshift ($z < 1.5$) dataset
with robust colours and galaxy shape information, and is not currently
optimised for high redshift and/or noisy data that is photometric
only, as was the case with the PHAT datasets.  However, it was useful
to examine its unoptimised performance with the PHAT data, as an
indication of the extent to which optimisation of the network
characteristics to a given input data scheme matters.

\section{PHAT1 - a test on GOODS data}
\label{sec:PHAT1}

\subsection{Motivation}
The estimation of photo-$z$'s is special in the sense that the desired
answer can in principle be obtained through spectroscopic
observations. Thus, we have an accurate benchmark which we can compare
photo-$z$'s to and we do not have to rely fully on simulations. This
is a very different situation from other estimation problems in
astronomy, e.g. the estimation of shapes of galaxies for weak
gravitational lensing, where accurate knowledge of the intrinsic shape
is inaccessible for comparison.

Given the high complexity of the photo-$z$ approach and the multiple
factors that influence the results it is reasonable to test the
photo-$z$ codes on real photometric data of objects that have also
been observed spectroscopically for precise redshift measurements. In
this way the tendency of simulations to idealise certain aspects of
real data can be avoided.

As a note of caution it should, however, be mentioned that comparisons
of photo-$z$'s to spec-$z$'s might well draw a somewhat idealised
picture of photo-$z$ performance. The currently available
spectroscopic catalogues are only highly complete at bright
magnitudes. For fainter magnitudes the fraction of high-quality
spectroscopic redshift measurements decreases. As
\cite{2008A&A...480..703H} showed, the objects missing in the spec-$z$
catalogues are likely the ones for which also photo-$z$ estimation is
harder and photo-$z$ accuracy is worse. We chose the GOODS-N field
also for the reason that it is one of the regions of the sky with the
most complete spectroscopy down to faint limits.

\subsection{Data set}
\label{sec:PHAT1_data_set}
The imaging data for this test are part of the Great Observatories
Origins Deep Survey northern field \citep[GOODS-N,
][]{2004ApJ...600L..93G}. The original four-band, optical ACS data are
complemented with images at other wavelengths from a variety of
instruments. See Table~\ref{tab:goods_data} for a summary. In total,
there are data in 18 bands covering the near-UV to the mid-IR.

The photometry used in the PHAT1 test is drawn from
\cite{2004AJ....127..180C} which includes $U$, $B_J$, $V_J$, $R_C$,
$I_C$, $z^\prime$ and $HK^\prime$ photometry.  Deep $J$, and $H$ band
photometry taken with ULBCAM on the UH2.2m \citep{2006ApJ...647...74W}
and $K_s$ band photometry taken with WIRC on Palomar
\citep{2005ApJ...625..621B} were added by first PSF matching then
measuring photometry in $3\arcsec$ diameter apertures using the method
described in \cite{2004AJ....127..180C}.  The GOODS-ACS photometry in
F435W (B), F606W (V+R), F775W ($i^\prime$), and F850LP ($z^\prime$)
along with the IRAC data (Moustakas et al. private Communication) were
added by positionally matching the catalogues provided by the GOODS
team with the \cite{2004AJ....127..180C} catalogues using a $1\arcsec$
matching radius.  Following recommended practice, the SExtractor
MAG\_AUTO magnitudes were used for the ACS data, while the aperture
corrected $3.6\arcsec$ diameter aperture magnitudes were used for
IRAC.

For this stage of testing we wanted to use publicly available data
that could be obtained with minimal effort by an average
researcher. The results of this test illustrate the critical role that
photometric methods play in obtaining good photo-$z$'s.  We strongly
recommend care in obtaining photometry across images with variable and
very different PSFs. Images should be aligned, the PSFs matched, and
fluxes measured in consistent apertures and care should be taken to
ensure noise estimates are correct
\citep{2004AJ....127..180C,2007ApJS..172...99C,2004A&A...421..913W,2001ApJS..135...41F}.
As illustrated by our test on one of the best studied fields in the
sky, correctly measured pan-chromatic photometry is not generally
available.  Users will likely have to, and probably should, measure
their own photometry to ensure the best results.  This is made simpler
by automated tasks such as ColorPro \citep{2006AJ....132..926C} which
measure PSF matched aperture photometry for a combination of space and
ground based data, while more complicated routines such as TFIT
\citep{2007PASP..119.1325L} fit high resolution galaxy images using
the local PSF for each image.

\begin{figure}
\resizebox{\hsize}{!}{\includegraphics[width=\textwidth]{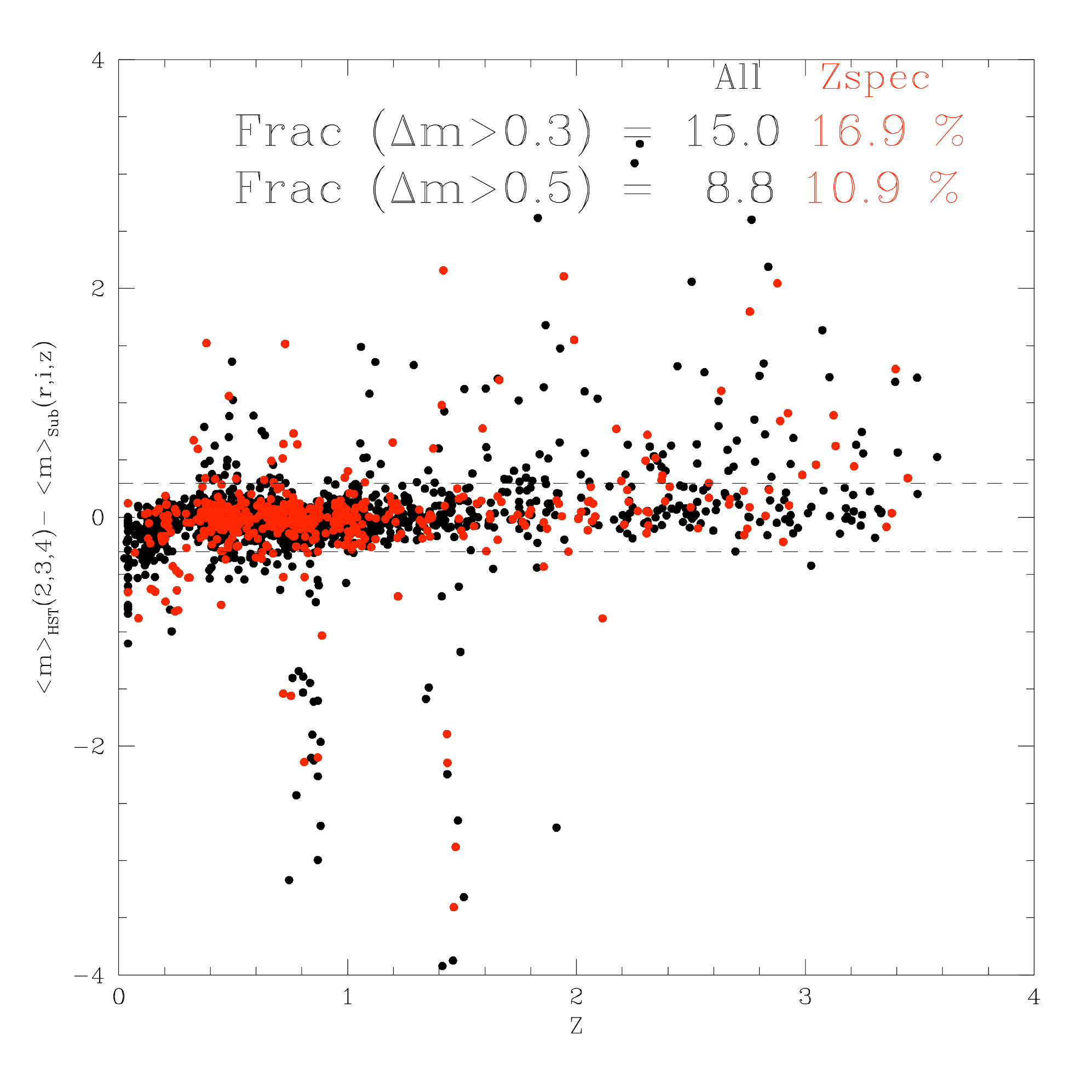}}
\caption{Difference between the average ACS (mean of F606W, F775W, and
  F850LP) and average SUPRIMECAM (mean of $RIz$) magnitudes as a
  function of redshift in the PHAT1 catalogue.}
\label{fig:discrepancy}
\end{figure}

Bulk photometric offsets were removed by minimising the offset between
the predicted and measured photometric points as a function of rest
frame magnitude as described in
\cite{2007ApJS..172...99C}.\footnote{Note that this procedure is only
  mildly dependent on the \cite{2007ApJS..172...99C} template set used
  for the re-calibration because the redshift range of the training
  sample is broad. For a given template SED the same rest frame
  wavelength corresponds to many different observer's frame
  wavelengths so that systematic features in a template get
  distributed evenly over many filters. Only BP-t, HY-t, and KR-t use
  template sets that are somewhat similar to the
  \cite{2007ApJS..172...99C} template set.}  The resulting photometry
has mean systematic offsets between photometric bands smaller than
0.01 mag. However, close inspection of the photometric catalogue shows
that there is a fraction of objects which show a rather large
discrepancy between the ACS- and the SUPRIMECAM-photometry in the
optical. Those objects are essentially evenly distributed in
redshift. A fraction of 15\% (10\%) of the objects shows a difference
of $>0.3{\rm mag}$ ($>0.5{\rm mag}$) between an object's average ACS
magnitude (mean of F606W, F775W, and F850LP) and an average SUPRIMECAM
magnitude (mean of $RIz$), as displayed in
Fig.~\ref{fig:discrepancy}. Some of these objects might be variable,
while others might be affected by different blending in the space- and
ground-based bands. We do not filter these objects because they are
also included in photometric catalogues that are routinely used for
many science projects. We want to provide estimates of photo-$z$
accuracy that are as close to reality as possible and such mismatches
of photometry from different instruments (or also different bands of
the same instrument) are not exceptions but rather the norm. Such
issues reflect the complex problem of obtaining a good photometric
catalogue from multi-band imaging data taken with different cameras
and/or taken under different observing conditions. But we will comment
upon the impact of these objects on global photo-$z$ performance in
the following sections and mention some strategies to prune them.

\begin{table}
\begin{minipage}[t]{\columnwidth}
\caption{Filters used for the PHAT1 test.}
\label{tab:goods_data}
\centering
\renewcommand{\footnoterule}{}  % to avoid a line before footnotes
\begin{tabular}{llr}
\hline
\hline
Filter & Instrument & $m_{\rm lim.; AB}$\\
\hline
$U$     & MOSAIC@KPNO-4m       & 27.1\footnote{5-$\sigma$ in a circular aperture with a diameter of $3\arcsec$}\\
$B$     & SUPRIMECAM@Subaru    & 26.9$^a$\\
$V$     & SUPRIMECAM@Subaru    & 26.8$^a$\\
$R$     & SUPRIMECAM@Subaru    & 26.6$^a$\\
$I$     & SUPRIMECAM@Subaru    & 25.6$^a$\\
$Z$     & SUPRIMECAM@Subaru    & 25.4$^a$\\
F435W   & ACS@HST              & 27.8\footnote{10-$\sigma$ in a circular aperture with a diameter of $0\farcs2$}\\
F606W   & ACS@HST              & 27.8$^b$\\
F775W   & ACS@HST              & 27.1$^b$\\
F850LP  & ACS@HST              & 26.6$^b$\\
$J$     & ULBCAM@UH-2.2m       & 24.1\footnote{5-$\sigma$ for a point-source}\\
$H$     & ULBCAM@UH-2.2m       & 23.1$^c$\\
$HK$    & QUIRC@UH-2.2m        & 22.1$^a$\\
$K$     & WIRC@Hale-5m         & 22.5\footnote{5-$\sigma$ for a Gaussian profile with FWHM=$1\farcs3$}\\
$3.6\mu{\rm m}$ & IRAC@Spitzer & 25.8\footnote{10-$\sigma$ for a point-source}\\
$4.5\mu{\rm m}$ & IRAC@Spitzer & 25.8$^e$\\
$5.8\mu{\rm m}$ & IRAC@Spitzer & 23.0$^e$\\
$8.0\mu{\rm m}$ & IRAC@Spitzer & 23.0$^e$\\
\hline
\end{tabular}
\end{minipage}
\end{table}

\begin{figure*}
\caption{$R$-band magnitude- (\emph{left}) and redshift-distributions
  (\emph{right}) of the PHAT1 catalogue (\emph{solid}) and the
  training sub-sample (\emph{dotted}).} \centering
\includegraphics[width=8cm]{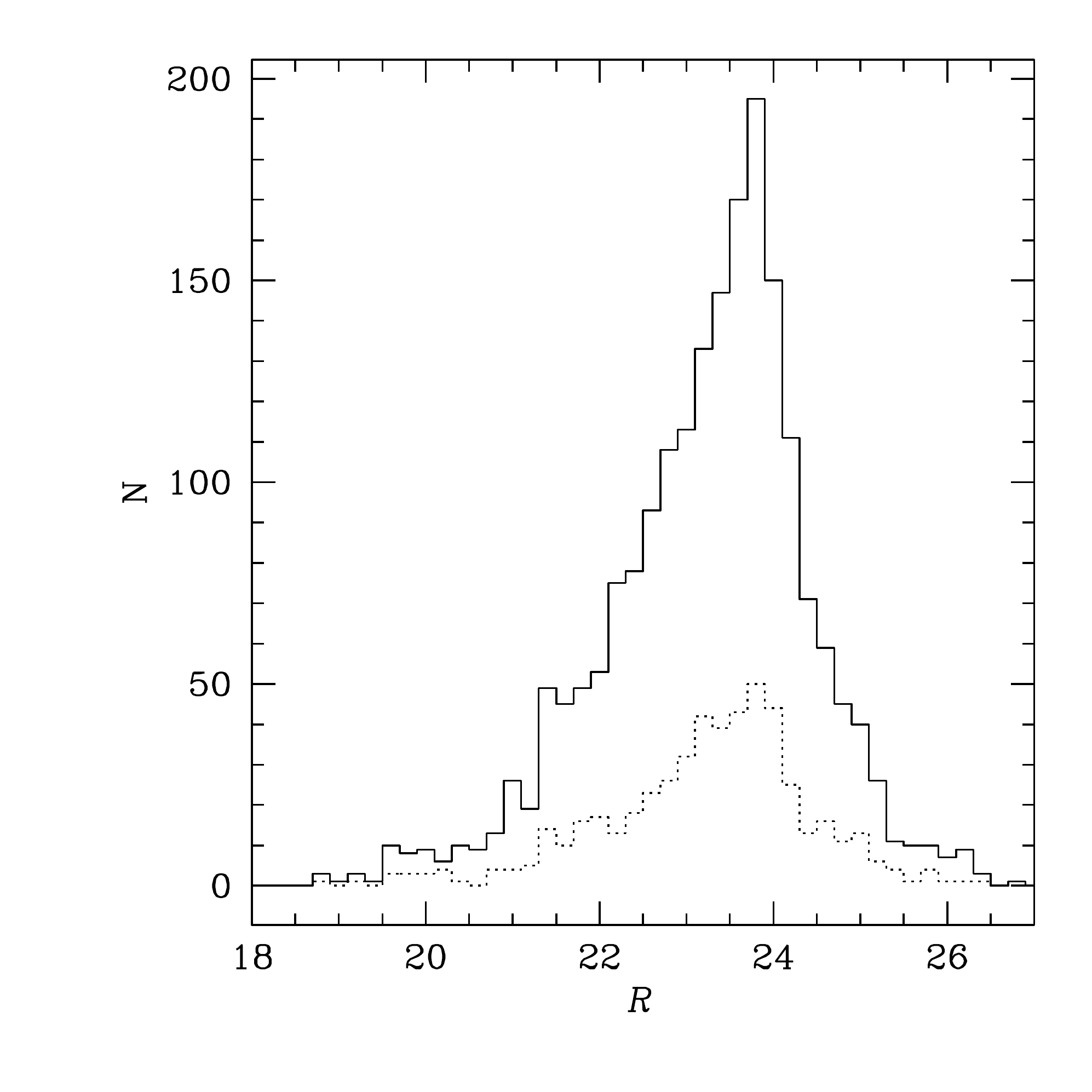}
\includegraphics[width=8cm]{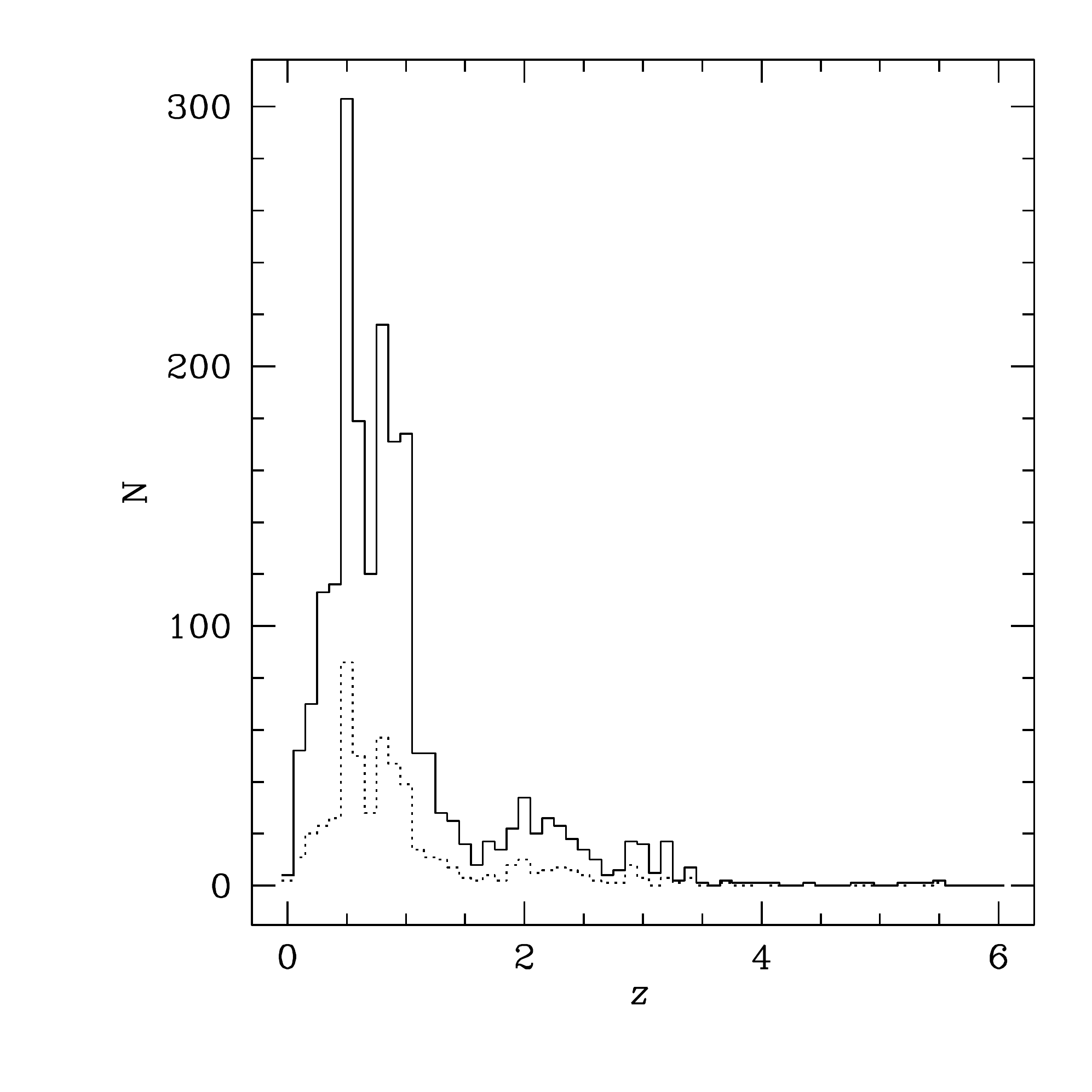}
\label{fig:mag_z_dist}
\end{figure*}

The photometric catalogue is matched to different spectroscopic
catalogues from \cite{2004AJ....127.3137C}\footnote{which includes
  spec-$z$'s from
  \cite{1996ApJ...471L...5C,2000ApJ...538...29C,2001AJ....121.2895C,1997ApJ...489..543P,1997ApJ...481..673L,1998ASPC..146..110L,1998hdf..symp..219D,1999AJ....118.1912L,2000AJ....119.2092B,2001AJ....122.2177B,2003AJ....126..632B,1996AJ....112..352S,2003ApJ...592..728S}},
\cite{2004AJ....127.3121W}, \cite{2005ApJ...622L...5T}, and
\cite{2006ApJ...653.1004R}\footnote{which includes spec-$z$'s from
  \cite{2004ApJ...611..725B}}.  This yields a total of 1984 objects
with 18-band photometry and spectroscopic redshifts. We randomly
select a quarter of those objects as a training set, i.e. for the
release of the catalogue the spectroscopic redshifts of one quarter of
the objects are revealed.\footnote{It should be noted that this is a
  fairly small training set for such a large redshift range. It cannot
  be expected that empirical codes perform as well on such a data set
  as template-based codes. This should not be regarded as a deficiency
  in the codes but rather a deficiency in the data.} The magnitude and
redshift distributions are shown in Fig.~\ref{fig:mag_z_dist}. Note
that the catalogue is highly complete down to $R\sim24$. The PHAT1
catalogue does not only contain normal galaxies. There is a small
number of AGN in the sample which we explicitly decided to include.

The participants are asked to run their codes twice on the provided
catalogue, once including the IRAC bands and once without the IRAC
bands. This is done because many template sets are inaccurate in the
mid-IR and we do not want this effect to dominate the
comparisons. Unlike in PHAT0 the participants using template-based
codes were asked to choose the best possible template set for their
code in PHAT1. Thus, template sets differ between the different ``-t''
methods here.

\subsection{Results for the 14-band case}
\label{sec:res14}
We use a similar set of statistics as for the PHAT0 test to
characterise the performance of the photo-$z$'s on the PHAT1 data with
two differences:
\begin{itemize}
\item We report the bias and scatter of $\Delta z' = \frac{z_{\rm
    spec}-z_{\rm phot}}{1+z_{\rm spec}}$.
\item Outliers are defined as objects with $|\Delta z'|>0.15$.
\end{itemize}

\begin{figure*}
\caption{Scatter and outlier values for the 14- (crosses) and 18-band
  (squares) PHAT1 case. The arrows indicate the effect of adding the
  IRAC bands on photo-$z$ accuracy. The \emph{left} panel shows the
  statistics for all objects and the \emph{right} panel the ones for
  all objects with an $I$-band magnitude $R<24$.} \centering
\includegraphics[width=8cm]{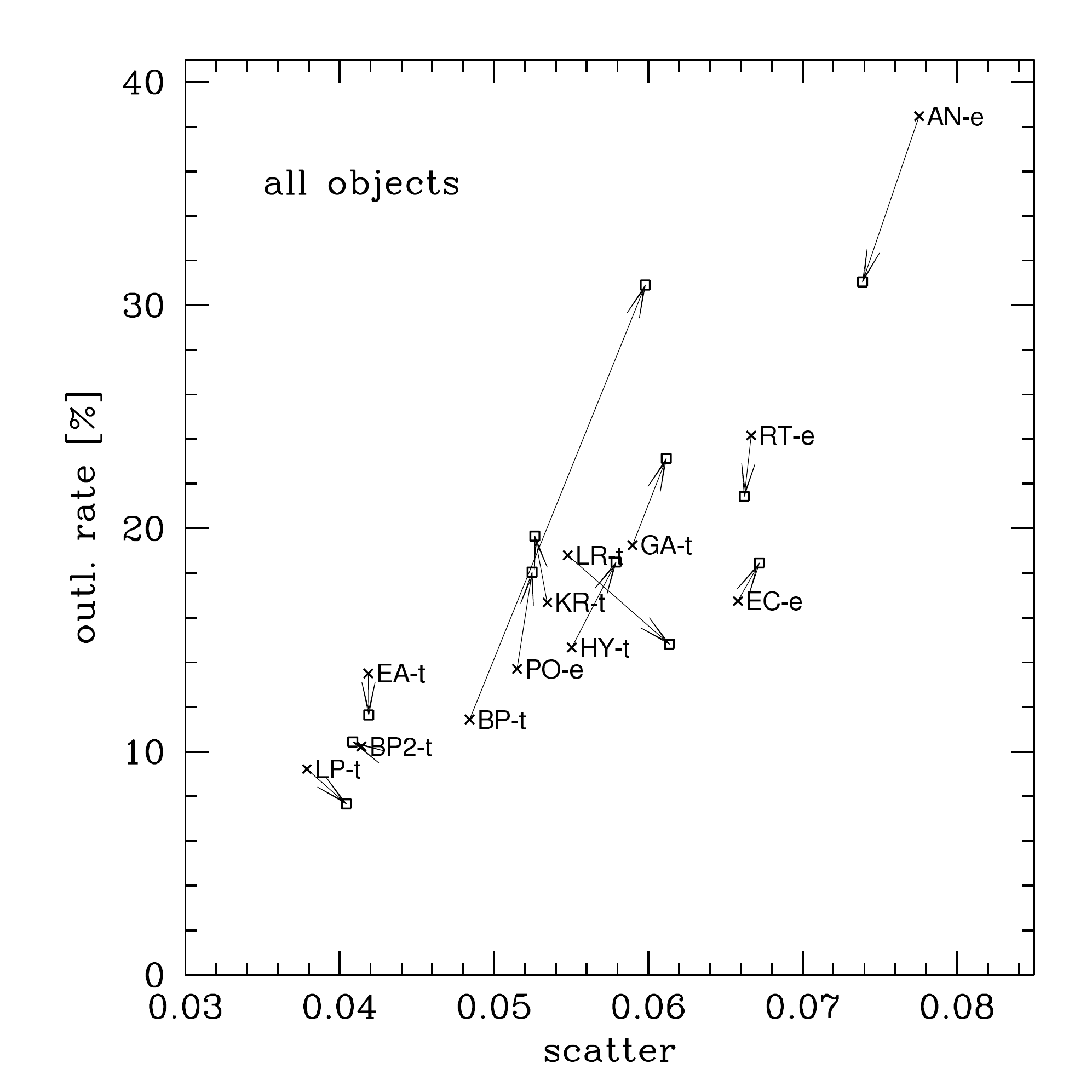}
\includegraphics[width=8cm]{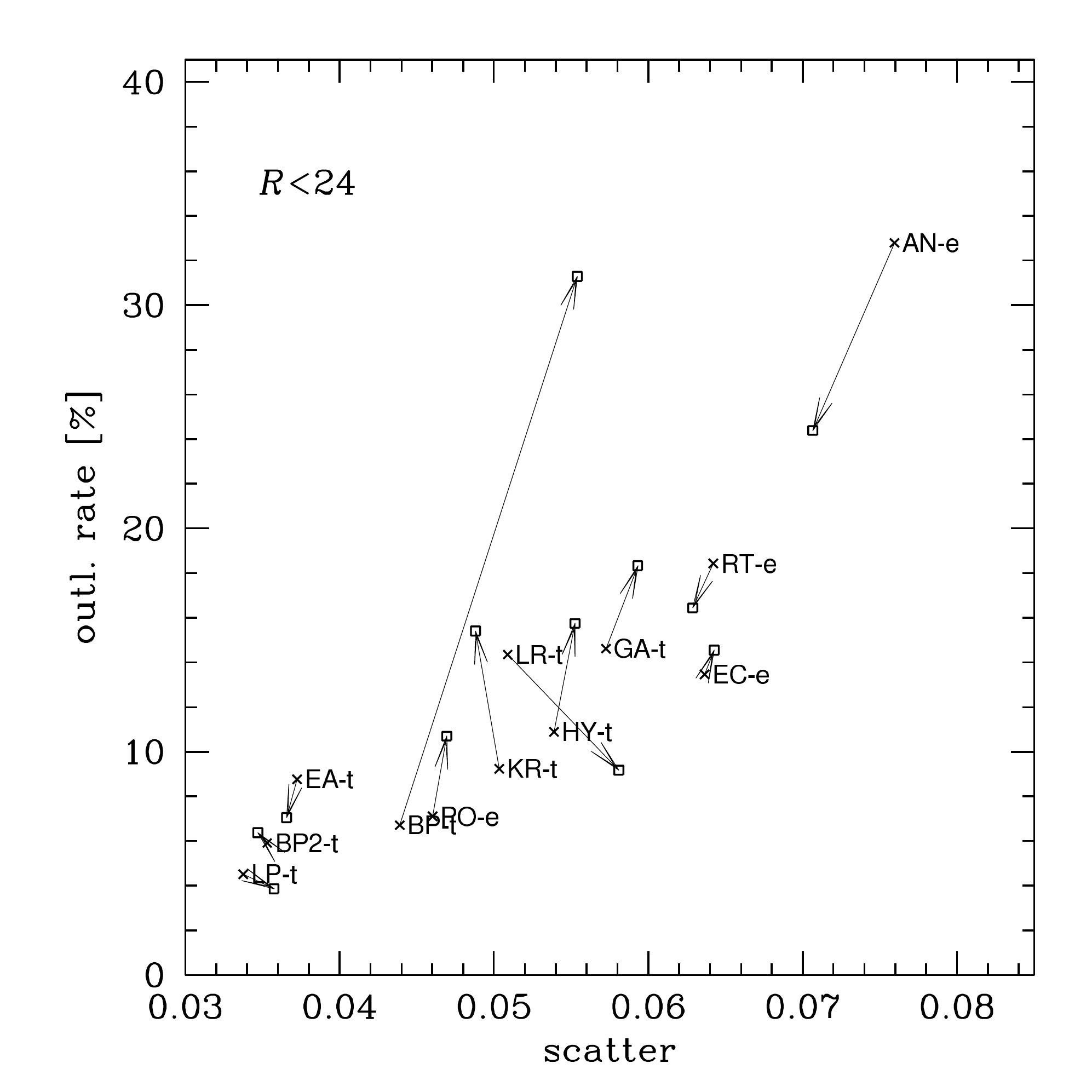}
\label{fig:stats_1418}
\end{figure*}

\begin{figure*}
\caption{Results of the PHAT1 test with 14 bands (i.e. excluding IRAC bands),  $z_{\rm phot}$ vs. $z_{\rm spec.}$. Objects with $R\ge 24$ are labelled in red.}
\centering
\includegraphics[width=17cm]{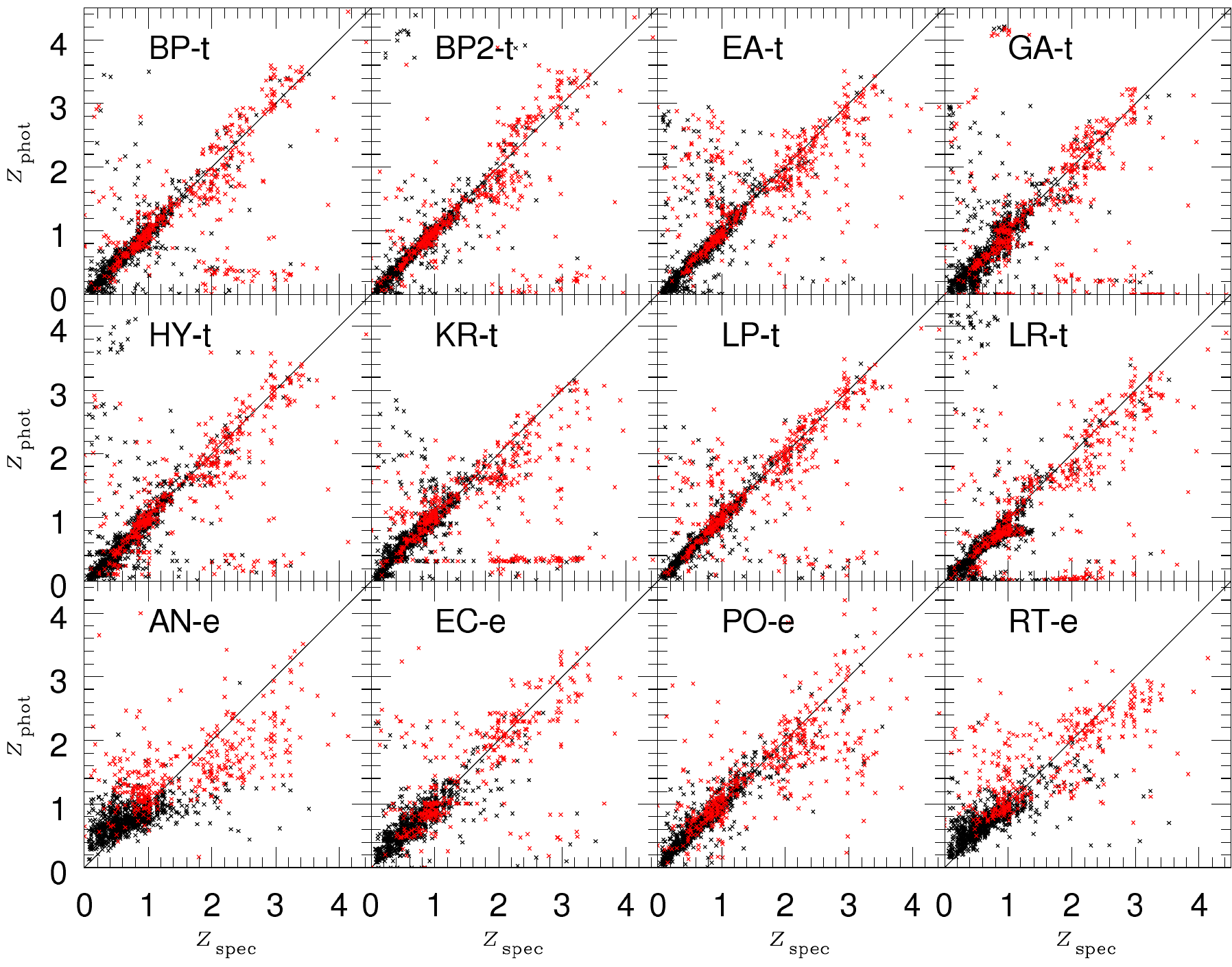}
\label{fig:res_HDFN_wo_sp}
\end{figure*}

\begin{figure*}
\caption{Similar to Fig.~\ref{fig:res_HDFN_wo_sp} but showing $\Delta z=z_{\rm spec.}-z_{\rm phot}$ vs. $z_{\rm spec.}$.}
\centering
\includegraphics[width=17cm]{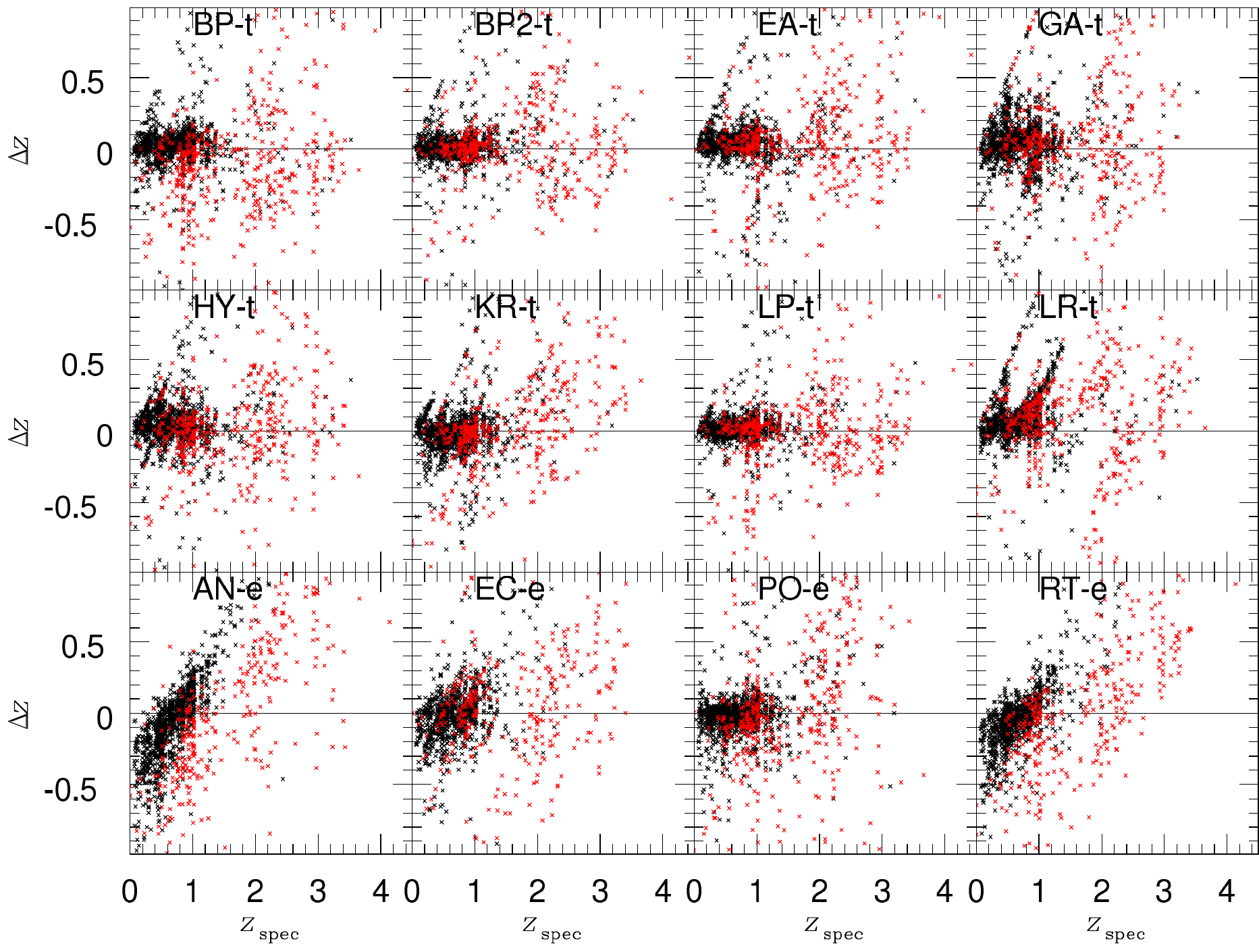}
\label{fig:res_HDFN_diff_wo_sp}
\end{figure*}

The resulting statistics are summarised in
Table~\ref{tab:res_goods}-\ref{tab:res_goods_z15}\footnote{In
  Table~\ref{tab:res_goods_05} results are presented for a relaxed
  definition of outliers being objects with $|\Delta z'|>0.5$.} and
the scatter and outlier values are plotted in
Fig.~\ref{fig:stats_1418} for the full sample and for an $R<24$
magnitude-limited sample. The full error distributions are displayed
in Figs.~\ref{fig:res_HDFN_wo_sp}~\&~\ref{fig:res_HDFN_diff_wo_sp} for
the 14-band case (i.e. without the IRAC bands). The results for the
empirical codes only include the non-training objects whereas the
results for the template-based codes include all objects. We checked
the performance of the template-based codes on the training and
non-training sample and found no significant differences.

\begin{table*}
\begin{minipage}[t]{\columnwidth}
\caption{Results for the PHAT1 catalogue with and without the IRAC
  bands, and for all objects and a magnitude-limited sample with
  $R<24$.}
\label{tab:res_goods}
\centering
\renewcommand{\footnoterule}{}  % to avoid a line before footnotes
\begin{tabular}{l|rrr|rrr|rrr|rrr}
\hline
\hline
        & \multicolumn{3}{|c|}{18-band} & \multicolumn{3}{|c}{14-band} & \multicolumn{3}{|c|}{18-band $R<24$} & \multicolumn{3}{|c}{14-band $R<24$}\\
Code & bias & scatter & outl.\footnote{Percentage of objects with \mbox{$|\Delta z'|=|\frac{z_{\rm spec}-z_{\rm phot}}{1+z_{\rm spec}}|>0.15$.} The numbers for the cleaned sample excluding objects with discrepant ACS/SUPRIMECAM photometry are given in brackets.} & bias & scatter & outl.$^a$ & bias & scatter & outl.$^a$ & bias & scatter & outl.$^a$ \\
\hline
BP-t & -0.046 & 0.060 & 30.9 (27.7) & 0.011 & 0.048 & 11.4 (7.1) & -0.053 & 0.055 & 31.3 & 0.012 & 0.044 & 6.7\\
BP2-t & 0.003 & 0.041 & 10.4 (7.5) & 0.004 & 0.041 & 10.2 (7.8) & 0.003 & 0.035 & 6.4 & 0.005 & 0.035 & 5.9\\
EA-t & 0.020 & 0.042 & 11.6 (5.9) & 0.022 & 0.042 & 13.5 (7.1) & 0.021 & 0.037 & 7.0 & 0.023 & 0.037 & 8.8\\
GA-t & -0.009 & 0.061 & 23.1 (18.1) & 0.016 & 0.059 & 19.3 (15.5) & -0.012 & 0.059 & 18.3 & 0.018 & 0.057 & 14.6\\
HY-t & -0.001 & 0.058 & 18.5 (15.2) & 0.018 & 0.055 & 14.7 (10.1) & -0.002 & 0.055 & 15.7 & 0.019 & 0.054 & 10.9\\
KR-t & -0.008 & 0.053 & 19.7 (13.3) & -0.006 & 0.053 & 16.7 (9.8) & -0.010 & 0.049 & 15.4 & -0.008 & 0.050 & 9.2\\
LP-t & 0.004 & 0.040 & 7.7 (4.9) & 0.009 & 0.038 & 9.2 (4.7) & 0.005 & 0.036 & 3.9 & 0.009 & 0.034 & 4.5\\
LR-t & 0.024 & 0.061 & 14.8 (12.9) & 0.038 & 0.055 & 18.8 (15.9) & 0.021 & 0.058 & 9.2 & 0.039 & 0.051 & 14.4\\
\hline
AN-e & -0.010 & 0.074 & 31.0 (29.0) & -0.006 & 0.078 & 38.5 (36.5) & -0.013 & 0.071 & 24.4 & -0.007 & 0.076 & 32.8\\
EC-e & -0.001 & 0.067 & 18.4 (15.3) & 0.002 & 0.066 & 16.7 (13.3) & -0.006 & 0.064 & 14.5 & -0.003 & 0.064 & 13.5\\
PO-e & -0.009 & 0.052 & 18.0 (14.5) & -0.007 & 0.051 & 13.7 (9.4) & -0.009 & 0.047 & 10.7 & -0.008 & 0.046 & 7.1\\
RT-e & -0.009 & 0.066 & 21.4 (19.0) & -0.008 & 0.067 & 24.2 (21.6) & -0.012 & 0.063 & 16.4 & -0.012 & 0.064 & 18.4\\
\hline
\end{tabular}
\end{minipage}
\end{table*}

\begin{table*}
\begin{minipage}[t]{\columnwidth}
\caption{Same as Table~\ref{tab:res_goods} but with a relaxed
  criterion for outliers.}
\label{tab:res_goods_05}
\centering
\renewcommand{\footnoterule}{}  % to avoid a line before footnotes
\begin{tabular}{l|rrr|rrr|rrr|rrr}
\hline
\hline
        & \multicolumn{3}{|c|}{18-band} & \multicolumn{3}{|c}{14-band} & \multicolumn{3}{|c|}{18-band $R<24$} & \multicolumn{3}{|c}{14-band $R<24$}\\
Code & bias & scatter & outl.\footnote{Percentage of objects with \mbox{$|\Delta z'|=|\frac{z_{\rm spec}-z_{\rm phot}}{1+z_{\rm spec}}|>0.5$.} The numbers for the cleaned sample excluding objects with discrepant ACS/SUPRIMECAM photometry are given in brackets.} & bias & scatter & outl.$^a$ & bias & scatter & outl.$^a$ & bias & scatter & outl.$^a$ \\
\hline
BP-t & -0.084 & 0.122 & 5.9 (5.0) & 0.016 & 0.085 & 4.8 (5.0) & -0.098 & 0.112 & 5.8 & -0.098 & 0.112 & 5.8\\
BP2-t & 0.009 & 0.084 & 3.8 (2.4) & 0.011 & 0.081 & 3.6 (2.4) & 0.008 & 0.072 & 1.5 & 0.008 & 0.072 & 1.5\\
EA-t & 0.023 & 0.088 & 4.2 (2.0) & 0.026 & 0.092 & 5.5 (2.0) & 0.024 & 0.074 & 1.9 & 0.024 & 0.074 & 1.9\\
GA-t & -0.014 & 0.125 & 8.7 (5.9) & 0.030 & 0.106 & 7.7 (5.9) & -0.026 & 0.115 & 5.4 & -0.026 & 0.115 & 5.4\\
HY-t & -0.011 & 0.116 & 4.9 (4.2) & 0.027 & 0.098 & 4.8 (4.2) & -0.016 & 0.109 & 3.5 & -0.016 & 0.109 & 3.5\\
KR-t & -0.015 & 0.114 & 8.6 (5.9) & -0.003 & 0.105 & 6.9 (5.9) & -0.024 & 0.101 & 6.6 & -0.024 & 0.101 & 6.6\\
LP-t & 0.003 & 0.079 & 2.3 (1.4) & 0.011 & 0.079 & 3.7 (1.4) & 0.005 & 0.060 & 1.0 & 0.005 & 0.060 & 1.0\\
LR-t & 0.028 & 0.104 & 4.5 (4.0) & 0.054 & 0.098 & 7.6 (4.0) & 0.023 & 0.087 & 2.5 & 0.023 & 0.087 & 2.5\\
\hline
AN-e & -0.036 & 0.151 & 3.1 (2.4) & -0.035 & 0.173 & 4.2 (2.4) & -0.047 & 0.130 & 1.4 & -0.047 & 0.130 & 1.4\\
EC-e & -0.007 & 0.120 & 3.6 (3.1) & -0.003 & 0.114 & 3.6 (3.1) & -0.015 & 0.106 & 1.9 & -0.015 & 0.106 & 1.9\\
PO-e & -0.013 & 0.124 & 3.1 (2.3) & 0.001 & 0.107 & 2.3 (2.3) & -0.020 & 0.098 & 1.2 & -0.020 & 0.098 & 1.2\\
RT-e & -0.031 & 0.126 & 3.2 (2.8) & -0.028 & 0.137 & 3.6 (2.8) & -0.034 & 0.111 & 1.4 & -0.034 & 0.111 & 1.4\\
\hline
\end{tabular}
\end{minipage}
\end{table*}

\begin{table*}
\begin{minipage}[t]{\columnwidth}
\caption{Same as Table~\ref{tab:res_goods} but in two different redshift bins.}
\label{tab:res_goods_z15}
\centering
\renewcommand{\footnoterule}{}  % to avoid a line before footnotes
\begin{tabular}{l|rrr|rrr|rrr|rrr}
\hline
\hline
        & \multicolumn{3}{|c|}{18-band $z_{\rm spec}\le1.5$} & \multicolumn{3}{|c}{14-band $z_{\rm spec}\le1.5$} & \multicolumn{3}{|c|}{18-band  $z_{\rm spec}>1.5$} & \multicolumn{3}{|c}{14-band $z_{\rm spec}>1.5$}\\
Code & bias & scatter & outl.\footnote{Percentage of objects with \mbox{$|\Delta z'|=|\frac{z_{\rm spec}-z_{\rm phot}}{1+z_{\rm spec}}|>0.15$.} The numbers for the cleaned sample excluding objects with discrepant ACS/SUPRIMECAM photometry are given in brackets.} & bias & scatter & outl.$^a$ & bias & scatter & outl.$^a$ & bias & scatter & outl.$^a$ \\
\hline
BP-t & -0.050 & 0.055 & 31.4 (27.5) & 0.013 & 0.044 & 7.2 (4.1) & -0.019 & 0.074 & 28.0 (28.9) & -0.001 & 0.075 & 35.3 (27.5)\\
BP2-t & 0.003 & 0.035 & 6.8 (4.9) & 0.005 & 0.035 & 6.5 (4.5) & 0.001 & 0.071 & 30.7 (25.1) & 0.001 & 0.075 & 31.0 (31.3)\\
EA-t & 0.021 & 0.037 & 9.9 (3.9) & 0.022 & 0.038 & 11.9 (4.9) & 0.014 & 0.065 & 21.3 (19.9) & 0.024 & 0.062 & 22.7 (22.3)\\
GA-t & -0.010 & 0.060 & 19.7 (14.6) & 0.018 & 0.057 & 16.4 (12.9) & 0.003 & 0.071 & 42.7 (42.7) & 0.008 & 0.073 & 35.0 (34.1)\\
HY-t & -0.003 & 0.055 & 16.5 (12.9) & 0.018 & 0.054 & 12.3 (8.9) & 0.014 & 0.072 & 29.7 (30.8) & 0.021 & 0.062 & 28.0 (18.5)\\
KR-t & -0.012 & 0.047 & 16.8 (11.8) & -0.011 & 0.050 & 10.5 (6.1) & 0.026 & 0.072 & 35.7 (24.2) & 0.042 & 0.062 & 51.3 (36.0)\\
LP-t & 0.005 & 0.037 & 6.2 (3.2) & 0.008 & 0.034 & 6.8 (2.8) & 0.002 & 0.059 & 15.7 (16.6) & 0.014 & 0.057 & 23.0 (18.0)\\
LR-t & 0.023 & 0.059 & 10.1 (8.3) & 0.039 & 0.053 & 15.1 (12.0) & 0.028 & 0.079 & 41.3 (45.0) & 0.037 & 0.070 & 39.7 (43.1)\\
\hline
AN-e & -0.017 & 0.070 & 27.6 (25.5) & -0.010 & 0.076 & 33.6 (31.6) & 0.051 & 0.078 & 50.7 (53.2) & 0.045 & 0.077 & 66.4 (70.3)\\
EC-e & -0.003 & 0.065 & 16.1 (12.9) & -0.000 & 0.064 & 14.5 (11.4) & 0.015 & 0.077 & 32.3 (32.3) & 0.015 & 0.077 & 29.5 (26.6)\\
PO-e & -0.012 & 0.049 & 12.6 (9.6) & -0.011 & 0.047 & 9.4 (6.0) & 0.019 & 0.075 & 48.3 (48.3) & 0.026 & 0.074 & 37.7 (32.7)\\
RT-e & -0.016 & 0.062 & 19.6 (17.0) & -0.014 & 0.064 & 21.1 (18.6) & 0.040 & 0.072 & 31.8 (32.9) & 0.039 & 0.071 & 41.9 (42.4)\\
\hline
\end{tabular}
\end{minipage}
\end{table*}

The most striking feature in Fig.~\ref{fig:res_HDFN_wo_sp} and
Table~\ref{tab:res_goods} is the large fraction of outliers ($> 9\%$
of the total sample) with catastrophically wrong photo-$z$'s. This
fraction is higher than typical literature estimates. It should be
emphasised that some of the objects included here are unusual in the
sense that they have SEDs different from normal galaxies
(e.g. AGNs). A small fraction is also influenced by blending effects
in the ground-based bands or variability, so that there is a mismatch
between the ACS and the SUPRIMECAM optical photometry. There may also
be a very small number of objects with wrong spec-$z$'s. But the bulk
of the outliers are real. If we reject objects which have discrepant
photometry between ACS and SUPRIMECAM (see
Sect.~\ref{sec:PHAT1_data_set}) the outlier rates decrease
considerably as indicated by the values in brackets in
Table~\ref{tab:res_goods}. The bias is largely unaffected by this
filtering and the scatter values do not decrease by more than 10\%
(both not given in Table~\ref{tab:res_goods}). We also test the most
accurate code in PHAT1 (LP-t) without ACS photometry. The statistics
of the problematic objects do not improve significantly although
excluding ACS removes the discrepancy between overlapping optical
filters. This suggests that most of the outliers amongst these objects
are not just outliers because their photometry is corrupted, but
rather because it is intrinsically harder to estimate photo-$z$'s for
them. We leave the detailed characterisation of these peculiar objects
(their morphology, SEDs, remaining photometric issues, etc.) to a
future study.

A lot of codes seem to have problems with identifying correctly the
redshifts of objects from the \cite{2006ApJ...653.1004R} sample with
$1.5\la z\la 3$. We explicitly decided to include those objects in the
test in order not to artificially idealise the situation. PHAT was
conceived to give a realistic picture of what can be achieved with
today's techniques. Those outliers reported here are present in deep
photometric catalogues and it is a delicate task for every scientist
to remove those or account for their effect. The fact that literature
values of outlier rates are usually smaller reflects the difficulty of
a blind test, but it most probably also reflects that our combined
spec-$z$ catalogue, explicitly including objects from the so-called
``redshift-desert'', is more complete and representative than some
other commonly used catalogues. Especially at $R<24$ our spec-$z$
catalogue is highly complete, and also for this bright cut the outlier
rates are rather large for most codes (see Table~\ref{tab:res_goods}
and the right panel of Fig.~\ref{fig:stats_1418}).

There are means of identifying outliers (poor fits, broad
redshift-probability functions, etc.) and photometric catalogues can
often be cleaned (e.g. by extraction flags) to yield much lower
outlier rates. Depending on the science application such a filtering
can be more or less applicable. For example, we showed that rejecting
objects with problematic photometry can improve the situation
considerably. However, photo-$z$'s are often used in a rather blind
fashion without extensive checking (often due to a lack of spec-$z$
comparisons) and filtering. Some science applications also rely on
redshifts for \emph{all} objects not allowing for filtering. For those
kind of applications the raw numbers reported by PHAT1 in
Table~\ref{tab:res_goods} are more informative than the cleaned ones
given in brackets.

The best performance on this data set is achieved by the LP-t, BP2-t,
EA-t, and BP-t codes, with LP-t showing the smallest scatter and
outlier rates. The empirical PO-e code follows closely. While EA-t and
BP-t also performed nicely on the PHAT0 test with noise (LP-t was used
for the creation of the PHAT0 simulations), the good results for PO-e
came as a surprise because this code ranked next to last in the PHAT0
test with noise. The sparse training set of PHAT1 ($\sim500$ objects)
is apparently large enough to fully exploit the capabilities of PO-e
because there are not too many degrees of freedom involved here. In
contrast, the empirical AN-e code that was in the top group for PHAT0
fails basically on PHAT1. The training set of PHAT1 is too sparse to
train the neural network over this large redshift range. Neural
networks are generally very good at interpolating smooth
functions. However, the colour-redshift mapping of galaxies is highly
complex in many places. Furthermore, there are ambiguities \citep[also
  called colour-redshift degeneracies ][]{2000ApJ...536..571B} in a
catalogue spanning a large redshift range, i.e. objects with very
different redshifts and very similar colours. In general, neural
networks, as the one used in AN-e, are not prepared to deal with such
ambiguities since they only assign one output redshift to a particular
point in colour space.

The top group of five (LP-t, BP2-t, EA-t, BP-t, and PO-e) is followed
by HY-t, KR-t, LR-t, GA-t, EC-e, and RT-e in approximately this
order. HY-t, KR-t and LR-t show some more or less pronounced, peculiar
features with a number of objects being assigned very similar
photo-$z$'s (horizontal features in
Fig.~\ref{fig:res_HDFN_wo_sp}). These features certainly have a large
influence on the statistics and prevent those codes from performing as
well as the top group although their error distribution in the core
looks very similar. GA-t and EC-e show clearly a larger scatter in the
core of the error distribution. The distribution for EC-e is smoother
but with a larger width resulting in the largest scatter (excluding
AN-e).

It is obvious that the empirical codes produce biases that are smaller
by typically a factor of two compared to the template-based codes. The
data-model match is by construction better in the empirical case. A
mismatch in the template-based case can be due to both, slightly
inaccurate templates and slightly inaccurate photometry. It should be
noted that it is very hard to achieve a photometric cross-calibration
accuracy spanning the whole wavelength range from the UV to the
mid-IR. EC-e, which was designed with the goal of being as bias-free
as possible, shows by far the smallest bias indeed. The combination of
a machine-learning algorithm and the proper use of PDFs pays off here.

\subsection{Results for the 18-band case}
\begin{figure*}
\caption{Similar to Fig.~\ref{fig:res_HDFN_wo_sp} but for 18 bands (i.e. including IRAC bands).}
\centering
\includegraphics[width=17cm]{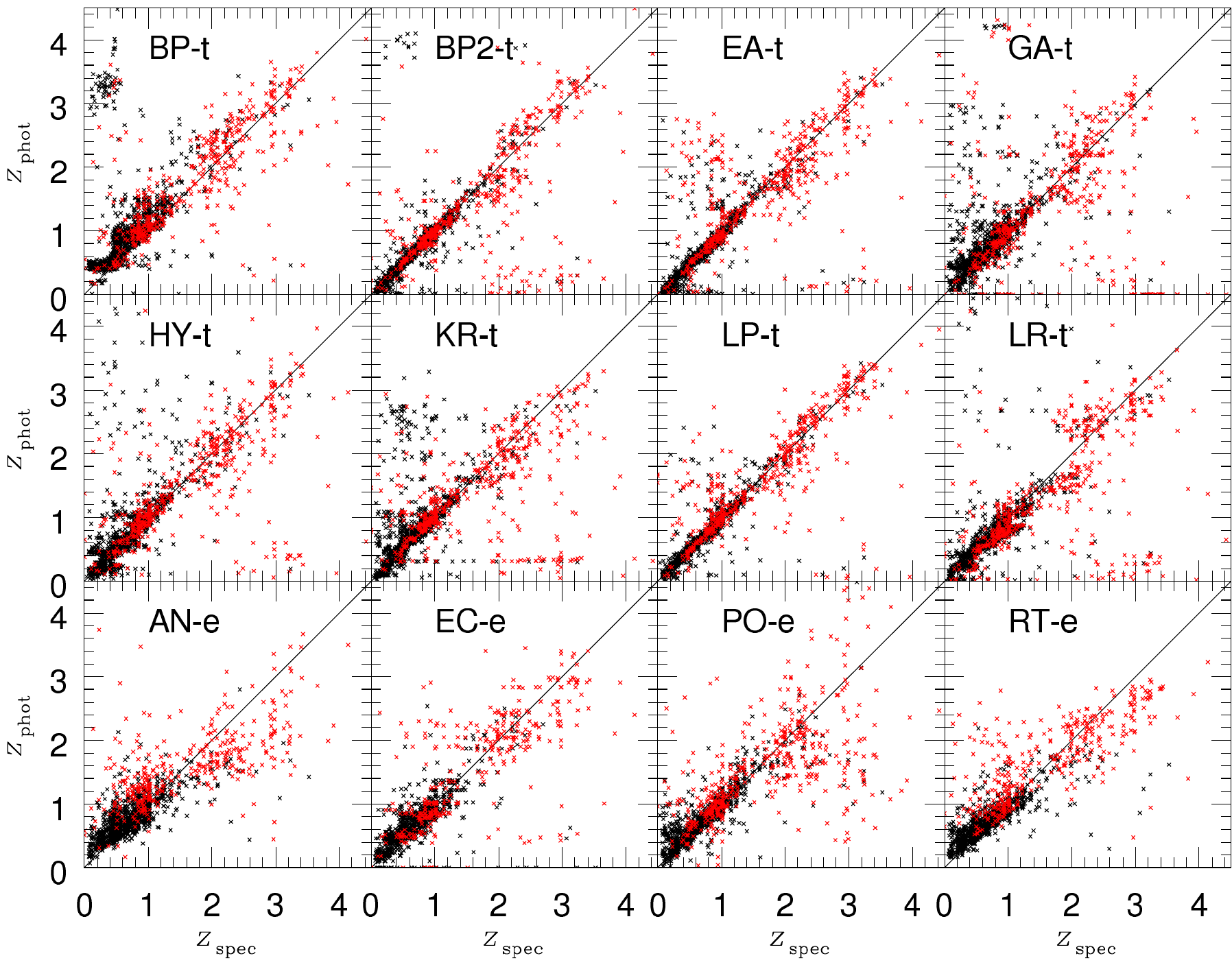}
\label{fig:res_HDFN}
\end{figure*}

\begin{figure*}
\caption{Similar to Fig.~\ref{fig:res_HDFN_diff_wo_sp} but for 18 bands (i.e. including IRAC bands).}
\centering
\includegraphics[width=17cm]{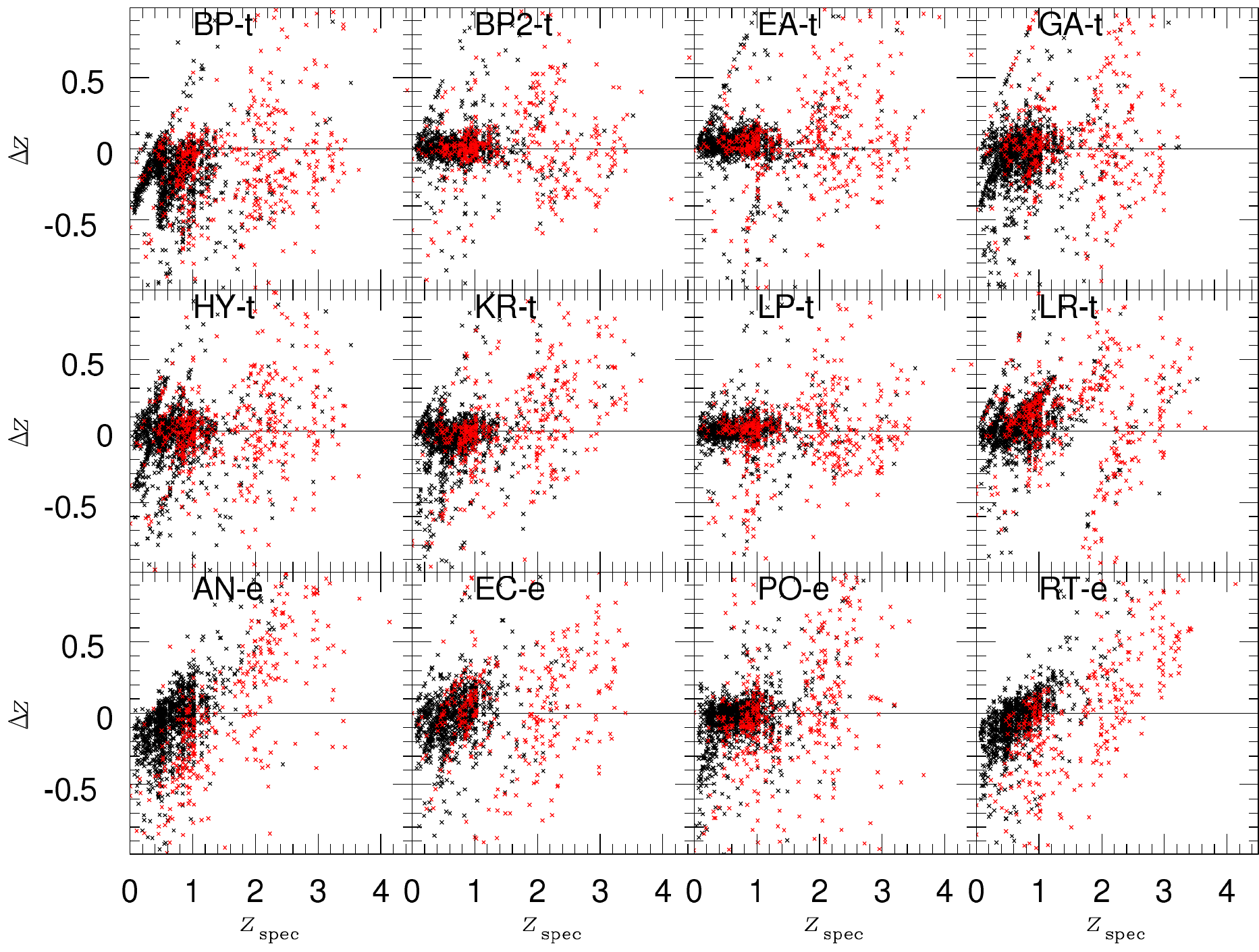}
\label{fig:res_HDFN_diff}
\end{figure*}

In Figs.~\ref{fig:res_HDFN}~\&~\ref{fig:res_HDFN_diff} the results for
the 18-band case (i.e. with IRAC bands included) are presented. The
statistics are also listed in Table~\ref{tab:res_goods} and the
scatter and outlier values for the different codes are plotted in
Fig.~\ref{fig:stats_1418} in comparison to the ones of the 14-band
case.

It is immediately obvious, especially from Fig.~\ref{fig:stats_1418},
that not all codes benefit from adding the IRAC photometry. Only LP-t,
EA-t, LR-t, RT-e, and AN-e show some improvement when adding those
information about the observed-frame mid-IR SEDs of the objects. The
outlier rates of LP-t and EA-t decrease by $\sim15\%$ compared to the
14-band case making them by far the best codes in this test, together
with BP2-t, which basically shows the same performance as with 14
bands. Also RT-e improves slightly in scatter and outlier rate with 18
bands compared to 14 bands. The bias and outlier rate of LR-t are
decreased somewhat but with the trade-off of a slightly larger
scatter. AN-e does not perform as poorly with 18 bands as with 14
bands but is still the least accurate code in this test.

PO-e, KR-t, HY-t, GA-t, and EC-e show slightly worse performance than
in the 14-band case with approximately conserved order. BP-t, however,
shows a huge increase in the number of outliers by $\sim200\%$ due to
a very poor low-$z$ performance. Most of the
\cite{2006AJ....132..926C} templates are undefined and must be
extrapolated for $\lambda > 25600\AA$.  These extrapolated SEDs have
significantly lower fluxes in the mid-IR compared to the observed IRAC
photometry resulting in a large photo-$z$ bias. Re-calibration of the
IRAC zeropoints with this template set improves the situation
somewhat, but is not done here for simplicity. The good performance of
BP2-t shows that it is not the code but the template set that makes
the difference here.

\subsection{Discussion of the PHAT1 results}
The performance shown by the best codes in the semi-blind PHAT1 test
with low bias and scatter values in the $4-5\%$ range is compatible
with typical literature values. Only the large fraction of outliers
($>7.5\%$) is worse than expected. We attribute this to the higher
completeness of our combined spec-$z$ catalogue besides the presence
of objects with unusual SEDs and some problems with the combination of
space-based and ground-based photometry. It should be noted that the
PHAT1 spectroscopic catalogue represents a very deep sample and is not
purely magnitude-limited. However, such depths are commonly used in
photometric studies in extragalactic astronomy. We cannot fully
quantify the fraction of outliers that are due to photometry problems
on the one hand or due to intrinsically problematic objects with
strange SEDs on the other hand. But the test of LP-t without ACS data
described in Sect.~\ref{sec:res14} suggests that most of the problem
seen here is connected to the latter.

Differences in the accuracy of the codes for the 14-band case can
mostly be attributed to differences in the template sets and priors
for the SED-fitting codes on the one hand and differences in the
training schemes for the empirical codes on the other hand. It is not
the aim of this study to explain all the features seen in this
comparison. Rather we want to provide a snapshot of what current codes
are capable to do in a semi-blind application.

It is striking that half of the codes perform worse with the IRAC
photometry included. Especially, the low-$z$ performance suffers in
this case. For the template-based codes this can be explained by
insufficient knowledge of the template SEDs in the
mid-IR\footnote{This is mostly due to insufficient modelling of dust
  emission features from PAHs.}. If the templates do not represent the
reality it cannot be expected that additional data lead to an
improvement. EA-t, the only template-based code that really benefits
from the information in the IRAC bands, differs from the other
template-based codes in the sense that it uses a template error
function \citep[see][for a detailed
  description]{2008ApJ...686.1503B}. This feature weighs the
measurements in the different bands according to the estimated
accuracy of the template at the rest-frame wavelength that corresponds
to the effective wavelength of a given filter at a particular redshift
step before computing the $\chi^2$. This hard-coded template error
function assigns a low accuracy to the mid-IR spectral region of the
templates so that the IRAC bands do not influence the $\chi^2$ at
low-$z$. At higher redshifts, however, when IRAC probes the rest-frame
near-IR or optical where templates are more accurate, the information
is used and can improve the photo-$z$'s. That is reflected in the
lower bias and outlier fraction for EA-t in the 18-band case when
compared to the 14-band case. BP2-t employs a filter error based on
the scatter between the photometry of best-fit models and observed
photometry in a particular filter on the spectroscopic training
set. This essentially down-weights the IRAC bands. In general the
mid-IR behaviour of the advanced template sets used by LP-t, BP2-t,
and EA-t seems to be more realistic than the extrapolations employed
for some other sets leading to better performance with 18 bands.

The lower bias values produced by the empirical codes suggest that
there are still systematic inaccuracies in most template sets. With a
sufficient training set such inaccuracies can be repaired by
re-calibrating the templates, e.g. with the approach described in
\cite{2000AJ....120.1588B}. Such a better data-model match is
demonstrated by BP2-t showing consistently the lowest bias of all
template-based methods which is however still somewhat larger than the
values for EC-e.

\section{Conclusions}
\label{sec:conclusions}
With PHAT we provide a snapshot of the photo-$z$ accuracy achievable
with today's methods in semi-blind tests. Most major photo-$z$ codes
used in the current literature are included in this challenge
presented here.

A first test, PHAT0, on highly idealised simulations yields good
agreement between the different codes (16 participants in total) and
especially in comparison to the LP-t code that was used
to create the simulations. Differences are found in the handling of
the opacity of the IGM, which are most likely unimportant for
practical applications (as long as only broad photometric bands are
used).

The PHAT1 test based on real photometric and spectroscopic data from
the GOODS survey represents a much more difficult test environment
including many of the challenges encountered in practical
applications. As expected the results from twelve participants show a
larger fluctuation in accuracy, but a general convergence is seen for
most codes, i.e. scatter values and outlier rates are within a factor
of two of the best code in the test. While the best codes perform to
expectations in terms of bias and scatter, some other codes show
remaining biases due to a template set that does not perfectly fit the
data or due to an insufficient training set. Half of the codes do not
benefit from adding mid-IR photometry from the Spitzer Space
Telescope. This finding suggest strongly that there is considerable
inaccuracy in some of the template sets in the rest-frame mid-IR
region of the SEDs. The rather large outlier rates reported in this
test should be taken seriously since most of these problematic objects
are also present in purely magnitude-limited photometric samples, but
not necessarily in commonly used spec-$z$ catalogues, which are
incomplete at fainter magnitudes. Cleaning of the catalogues is still
necessary for PHAT1 to reach an outlier rate below $\sim5\%$ for the
best code in the test. More detailed future studies (possibly in the
framework of PHAT) are needed to identify the nature of this problem
and quantify the contributions from multi-colour photometry issues on
the one hand and objects with intrinsically unusual SEDs on the other
hand. We believe that solving the problem of these outliers lies at
the core of future photo-$z$ improvements. It is clear that improved
spec-$z$ catalogues which are as complete as possible will be
indispensable for such an effort. Some science applications that do
not rely on complete samples of galaxies (like e.g. dark energy
studies with weak gravitational shear) can greatly benefit from
efficient cleaning of galaxy catalogues. There are ways of
considerably improving photo-$z$ accuracy by rejecting objects with
unreliable estimates. It is, however, beyond the scope of this study
to present strategies on how to optimise catalogues for different
science applications and how to quantify those improvements.

Photo-$z$ accuracy is of paramount importance for a large number of
future science projects, ranging from galaxy evolution to
cosmology. The differences in the performance of the different
photo-$z$ codes presented here will have a direct impact on the power
of photometric surveys to answer those scientific questions. We did
not quantify the impact of photo-$z$ accuracy here, but it should be
noted that there is still some way to go before photo-$z$'s reach the
accuracy required for e.g. future full-sky dark energy surveys.

The test environments used in this study are publicly available at
\url{http://www.astro.caltech.edu/twiki_phat/bin/view/Main/WebHome}
and can be used to assess the performance of future methods in
comparison to the results presented here in a quantitative and
unbiased way.

\begin{acknowledgements}
We would like to thank JPL/Caltech for hospitality and support during
the 2008 PHAT workshop. We are grateful to the large number of
colleagues who made PHAT a success through discussions, criticism, and
encouragement. A special thanks goes to Mike Hudson who came up with
the acronym ``PHAT''. HH would like to thank in particular Catherine
Heymans, Konrad Kuijken, Ludovic van Waerbeke, and Peter Schneider for
supporting the PHAT effort. HH was supported by the European DUEL RTN,
project MRTN-CT-2006-036133. The work of LAM and DC was carried out at
the Jet Propulsion Laboratory, California Institute of Technology,
under a contract with NASA. LAM acknowledges support by the NASA ATFP
program. CW was supported by an STFC Advanced Fellowship. NP
acknowledges support from NKTH:Polanyi and KCKHA005 grants.
\end{acknowledgements}

\bibliographystyle{aa}

\bibliography{PHAT}

\end{document}